%% file: main.tex
\documentclass[fleqn,usenatbib]{mnras}

\usepackage{newtxtext,newtxmath}
\usepackage[T1]{fontenc}

\DeclareRobustCommand{\VAN}[3]{#2}
\let\VANthebibliography\thebibliography
\def\thebibliography{\DeclareRobustCommand{\VAN}[3]{##3}\VANthebibliography}

\usepackage{graphicx}	% Including figure files
\usepackage{amsmath}	% Advanced maths commands
\usepackage{longtable}  % longtable
\usepackage{multirow}
\usepackage{manyfoot}
\usepackage{lipsum}
\usepackage{mathtools}
\usepackage{mathrsfs}
\usepackage{placeins}
\usepackage{booktabs}
\usepackage{xcolor}
\usepackage{deluxetable}
\usepackage{lscape}
\usepackage{rotating}
\usepackage{multirow}
\usepackage{graphicx}
\usepackage[normalem]{ulem}

\newcommand{\msol}{M$_\odot$}
\newcommand{\J}{XTE~J1814$-$338}
\newcommand{\vr}{\vspace{1mm}}
\definecolor{darkgreen}{rgb}{0.18, 0.55, 0.34}
\definecolor{orange}{rgb}{1.0, 0.27,0.0}
\definecolor{mygreen}{rgb}{0,0.502,0}

\title[Constraining XTE J1814-338 Properties via PPM]{Constraining the Properties of the Thermonuclear Burst Oscillation Source XTE J1814-338 Through Pulse Profile Modelling}

\author[Kini~et~al.]{Yves~Kini$^{1}$
\thanks{E-mail: \href{mailto:y.kini@uva.nl}{y.kini@uva.nl}},
Tuomo~Salmi$^{1}$, 
Serena~Vinciguerra$^{1}$,
Anna~L.~Watts$^{1}$,
Anna Bilous$^{2}$,\newauthor  
Duncan K. Galloway$^{3}$,
Emma van der Wateren$^{4}$,
Guru Partap Khalsa$^{1}$, 
Slavko~Bogdanov$^{5}$, \newauthor  
Johannes Buchner$^{6}$,
Valery~Suleimanov$^{7}$ 
\\
$^{1}$Anton Pannekoek Institute for Astronomy, University of Amsterdam, Science Park 904, 1090GE Amsterdam, the Netherlands\\
$^{2}$Independent researcher\\
$^{3}$School of Physics \& Astronomy, Monash University, Clayton, VIC 3800, Australia\\
$^{4}$ASTRON, Netherlands Institute for Radio Astronomy, Oude Hoogeveensedijk 4, 7991 PD, Dwingeloo, The Netherlands \\
$^{5}$Columbia Astrophysics Laboratory, Columbia University, 550 West 120th Street, New York, NY 10027, USA\\
$^{6}$Max Planck Institute for Extraterrestrial Physics, Giessenbachstrasse, 85741 Garching, Germany\\
$^{7}$Institut f\"ur Astronomie und Astrophysik, Kepler Center for Astro and Particle Physics, Universit\"at T\"ubingen, Sand 1, D-72076 T\"ubingen, Germany
}

\date{Accepted XXX. Received YYY; in original form ZZZ}

\pubyear{20XX}

\begin{document}
\label{firstpage}
\pagerange{\pageref{firstpage}--\pageref{lastpage}}
\maketitle

\input{abstract}


%%%%%%%%%%%%%%%%%%%%%%%%%%%%%%%%%%%%%%%%%%%%%%%%%%

%%%%%%%%%%%%%%%%% BODY OF PAPER %%%%%%%%%%%%%%%%%%
\input{Sec0-Introduction}
\input{Sec1-Modeling}

\input{Sec2-Results}

\input{Sec3-Discussion}

\section*{Acknowledgements}

YK, TS, SV, and ALW acknowledge support from ERC Consolidator Grant No.~865768 AEONS (PI Watts). We thank Jan-Erik Christian for the discussion on hybrid stars. This work was carried out on the HELIOS cluster, mostly on dedicated nodes funded via the abovementioned ERC CoG. EvdW contributed to this work as part of her Masters thesis project, and GPK as part of his Bachelor Project, both at UvA. SB acknowledges support from NASA grant 80NSSC22K0728. VS acknowledges support by Deutsche Forschungsgemeinschaft (DFG; grant WE 1312/59-1).

\section*{Data availability}
All the data, posterior samples as well as the scripts used for the runs and plots are available on Zenodo (\url{https://doi.org/10.5281/zenodo.8365643}).



\bibliographystyle{mnras}
\bibliography{bibliography}

%%%%%%%%%%%%%%%%%%%%%%%%%%%%%%%%%%%%%%%%%%%%%%%%%%

%%%%%%%%%%%%%%%%% APPENDICES %%%%%%%%%%%%%%%%%%%%%

\appendix
\input{Sec4-Appendix.tex}

\FloatBarrier

%\input{Comments}

%\newpage
%\input{Sec5-Comments}



%%%%%%%%%%%%%%%%%%%%%%%%%%%%%%%%%%%%%%%%%%%%%%%%%%

% Don't change these lines
\bsp	% typesetting comment
\label{lastpage}
\end{document}

%% file: abstract.tex
\begin{abstract}
Pulse profile modelling (PPM) is a comprehensive relativistic ray-tracing technique employed to determine the properties of neutron stars. In this study, we apply this technique to the Type I X-ray burster and accretion-powered millisecond pulsar \J, extracting its fundamental properties using PPM of its thermonuclear burst oscillations. Using data from its 2003 outburst, and a single uniform temperature hot spot model, we infer \J\ to be located at a distance of $7.2^{+0.3}_{-0.4}$ kpc, with a mass of $1.21^{+0.05}_{-0.05}$ \msol ~and an equatorial radius of $7.0^{+0.4}_{-0.4}$ km. Our results also offer insight into the time evolution of the hot spot but point to some potential shortcomings of the single uniform temperature hot spot model. We explore the implications of this result, including what we can learn about thermonuclear burst oscillation mechanisms and the importance of modelling the accretion contribution to the emission during the burst.  
\end{abstract}

%% file: Sec0-Introduction.tex
\section{Introduction}\label{sec:intro}

Accreting Millisecond X-ray Pulsars (AMXPs) involve a Neutron Star (NS) accreting material, mainly  hydrogen and/or helium,  from either a degenerate or a non-degenerate star (with mass $< 1$\msol) through a Roche lobe overflow \citep{Patruno21}. The relatively weak magnetic field of the NS in these systems ($\sim 10^7-10^9$ G) leads to the accreted material being channelled to the NS's magnetic poles through an accretion column. The in-falling plasma spreads across the NS's surface, forming a thick plasma shell. With the continuous influx of new material, the previously gathered plasma undergoes hydrostatic compression, leading to significant thermal energy buildup, ultimately culminating in ignition and nuclear burning \citep[see e.g.][for ignition conditions and burning regime]{Bildsten:1998, Keek:2016}. When the rate of nuclear burning exceeds the cooling rate, a thermonuclear explosion occurs on the surface of the NS. These thermonuclear explosions, commonly referred to as Type I X-ray bursts, emit intense X-ray radiation \citep[see ][for recent reviews]{Galloway:2020, Alizai:2023}.

The discovery of coherent pulsations \citep{Strohmayer:1996}, commonly known as thermonuclear burst oscillations, in certain X-ray bursts from specific sources has sparked significant interest in investigating these sources. This interest is motivated not only by the desire to understand the underlying causes of the burst oscillations \citep[see e.g.][]{Spitkovsky:2002,Heyl:2004, Lee:2004, Piro:2005,Cavecchi:2013,Cavecchi:2015, Cavecchi:2016, Mahmoodifar:2016, Chambers:2019, Garcia:2018, Chambers:2020, vanBaal:2020} but also because these pulsations offer a means to infer important properties of NSs \citep{Bhattacharyya:2004pp, Lo:2013ava}. In particular, inferring the mass and radius can provide crucial insights into the equation of state for NSs \citep[][]{ Lattimer:2012, Oertel:2016, Baym:2017, Tolos:2020, Yang:2019, Hebeler:2020}.

Inferring stellar properties using burst oscillations has proven to be complex. This is primarily due to the uncertainties linked to the burning process and oscillation properties, which render modelling of the short time-scale variability associated with the burst and burst oscillations computationally demanding. \citet{kini:2023a} have shown that failing to accurately model these short time-scale variabilities results in biased estimates of mass and radius, particularly for the high number of counts that are needed to derive meaningful constraints on these parameters. To address this bias, bursts can be divided into shorter segments (where variability can be overlooked) while simultaneously fitting the segments \citep{kini:2023b}. This is referred to as the \textit{slicing method}. \citet{kini:2023b} also verified that posterior distributions of mass and radius, derived from several bursts originating from the same source, can be combined to yield results similar to that of a single burst with an equivalent total number of counts.   

\citet{kini:2023a, kini:2023b} employed the Pulse Profile Modelling (PPM) technique, which relies on relativistic ray-tracing. This technique leverages the fact that observed X-ray pulsations carry information about the source's intrinsic properties \citep[see e.g.:][]{Pechenick:1983,Chen:1989,Page:1995,Miller:1998,Braje:2000, Weinberg:2001,Beloborodov:2002, Poutanen:2006, Cadeau:2007,Morsink:2007, Baubock:2012,Lo:2013ava, Psaltis:2013fha, Miller:2015,Stevens:2016, Nattila:2018, Bogdanov:2019}. Both of our previous studies used the X-ray Pulse Simulation and Inference \cite[\texttt{X-PSI};][]{Riley2023} package, initially developed to model X-ray pulsations from Rotation-Powered Millisecond Pulsars (RMPs) observed by the Neutron Star Interior Composition Explorer (NICER, \citealt{NICER}, and see \citealt{Riley:2019, Riley:2021, Salmi:2022,Salmi:2023,Vinciguerra:2023,Choudhury:2024, Salmi:2024_submitted,Vinciguerra:2024} for the \texttt{X-PSI} NICER RMP papers). Despite uncertainties about the size, shape, and temperature distribution of the hot spots for RMPs, their pulsations are stable, implying that short-time-scale variability can be overlooked unlike in the case of burst oscillation sources. 

To obtain the NS mass and radius, \texttt{X-PSI} computes the expected X-ray pulse profile (phase and spectrally resolved count rate) for a given model and a parameter vector using the PPM technique. \texttt{X-PSI} then compares the expected (model) and observed (data) X-ray pulse profiles by computing the likelihood of the observed X-ray pulse profile given that specific parameter vector and model. Using a Bayesian framework, \texttt{X-PSI} not only derives the posterior probability distributions for space-time parameters (mass and radius) but also for the hot spot properties, which are part of the model. 

\citet{kini:2023a,kini:2023b} used synthetic data and phenomenological models to mimic the observed bursts and burst oscillation properties from  \J. \J \ was first observed in June 2003 during a scan of the Galactic center \citep{Markwardt:2003}, when it underwent its only detected outburst. It is an AMXP with a pulsation frequency of 314.4 Hz and an orbital period of 4.275 hours \citep{Markwardt:2003, Papitto2007}. During this singular outburst, 28 Type I X-ray bursts were detected over about 50 days, each lasting around 2 minutes. Notably, burst oscillations were observed in all 28 bursts \citep{Strohmayer:2003}. These oscillations were phase-locked to the persistent pulsations and remained remarkably stable throughout, with an average root mean square fractional amplitude (rms FA) of 10\% \cite[see e.g.][]{Watts:2005}. This stability persisted across all the bursts, except for the 28th burst (corresponding to Burst 27 in \citealt{Strohmayer:2003}), which showed marginal evidence of photospheric radius expansion (PRE). Constraints on XTE J1814-338 have been obtained through studies of its quiescent optical counterpart \citep{Baglio:2012}, X-ray and optical imaging of the optical counterpart \citep{Krauss:2005sj}, (candidate) PRE burst \citep{Strohmayer:2003, Galloway:2020}, and burst oscillations \citep[][and see Section \ref{sec:compare} for further discussion on parameter constraints]{Bhattacharyya:2004pp, Leahy:2009}. \J\ was previously identified as one of the most promising sources for PPM \citep{Bhattacharyya:2004pp}, and in this paper we apply the \textit{slicing method} outlined by \citet{kini:2023b} to the observational data of its non-PRE bursts (all bursts except Burst 28), and infer its mass and radius. We also aim to understand better the conditions and the behaviour of the hot spot during the thermonuclear bursts.

The paper is structured as follows. In Section \ref{sec:modelging}, we present the observations, the data reduction, and the instrument models used. This is followed by an explanation of the inference procedures and a summary of the approach we used to combine posterior distributions from multiple bursts to obtain tighter constraints on parameters shared across all bursts. We present the main findings in Section \ref{sec:result}, followed by an in-depth discussion of these findings in Section \ref{sec:discussion}. We summarize our conclusions in Section \ref{sec:conclusion}.

%% file: Sec1-Modeling.tex
\section{Modelling }\label{sec:modelging} 

In this section, we first provide an overview of the data employed for this analysis and explain the data reduction process. Next, we describe the procedure for extracting the response matrices corresponding to each burst. Finally, we outline the inference approach, focusing on the various model components and the assumptions made to infer the properties of \J .

\subsection{Burst observations and data reduction}\label{sec:data}

The 28 type I X-ray bursts detected over about 50 days \citep{Strohmayer:2003,Watts:2005,Galloway:2020} with the Rossi X-Ray Timing Explorer \citep[RXTE;][]{Jahoda:1996} Proportional Counter Array were recorded in Science Event Mode (except for Burst 1), making them suitable for pulse profile modelling. The data for all bursts, except for the first one were acquired with the time resolution of $2^{-13}\,\mathrm{s}\approx122\,\mu$s and 64 energy bins spanning the range between 2 and 106\,keV (mode  E\_125us\_64M\_0\_1s)\footnote{For an example of energy-channel conversion tables, see \url{https://heasarc.gsfc.nasa.gov/docs/xte/e-c_table.html}}. Burst 1 was captured in GoodXenon mode, with a time resolution of 1\,$\mu$s and 256 energy bins spanning the range between 0.4 and 108 keV. In this analysis, we only used channel subset [1,30), corresponding to the nominal photon energy ranging from about 2 keV to about 26 keV for all bursts except Burst 1. For Burst 1, we adjusted the channel subset to [4,30) due to lower counts in the channels below channel 4, corresponding to the nominal photon energy range [1.6, 12.28] keV. We generated individual response matrices for each burst using the \texttt{pcarsp} tool from the HEASoft package (version 6.28).

During the observations, four Proportional Counter Units (PCUs) were active. For bursts 5 and 25, PCU0 was 
excluded from analysis because of abnormally large count rates suggesting a detector breakdown.

The start of the burst was defined as the point where the count rate first exceeded the pre-burst mean plus five times the standard deviations. Similarly, the end of the burst was identified as the last instance where the count rate surpassed this threshold. The on-burst and baseline windows were determined using the light curve, binned in 0.2-second intervals. For the baseline window, we selected the first and last 100 seconds of the data, calculating the mean count rate for each segment. The segment with the lowest mean was designated as the pre-burst count rate reference. Burst start times and durations are given in Table~\ref{tab:data_table}.

For the on-burst window, photon arrival times were barycentered with \texttt{faxbary} (from the HEASoft package, version 6.28) using spacecraft orbit files and DE405 Solar system ephemerides, with the source coordinates taken from \citet{Papitto2007}. The same work provided spin frequency ($\approx314.4$\,Hz) and binary ephemerides for computing the spin phase of each photon. The accuracy of the spin phase calculation was verified by visual inspection of the waterfall plot made from the whole dataset. The frequency of burst oscillations is known to fluctuate around the frequency of accretion-powered pulsations used for spin phase calculations \citep{Cavecchi:2022}, with the average phase discrepancy reaching 0.04 of spin phase during burst peaks. This is smaller, but comparable to our chosen spin phase bin of 0.0625 (we binned the counts in 16 phase bins).

\subsection{Inference}\label{sec:inference}

\subsubsection{Models}\label{sec:models}

The models employed to infer \J's\ parameters in this work closely resemble those used in \citet{kini:2023b}. Here, we outline the key model components and highlight any areas where the approach differs.

In both \citet{kini:2023b} and this paper, we assume the star to be oblate, with the external space-time curvature described by the Schwarzschild metric \citep{Morsink:2007}. After computing the light bending, corrections for relativistic Doppler boosting and aberration are incorporated to account for stellar rotation. We use state-of-the-art neutron star atmospheres developed for thermonuclear burst sources \citep{Valery2012}, assuming a composition matching solar abundances \citep[as indicated by the burst properties,][]{Galloway:2008}. For interstellar medium (ISM) absorption, we adopt the neutral gas absorption model \texttt{Tbabs}, which employs the photoelectric absorption cross-section derived by \citet{Wilms:2000}. Regarding the X-ray background photons, we make two alternative assumptions described in Section \ref{sec:bkg}. The likelihood of each sample is computed by marginalization over the background \citep[See Appendix B.2 of ][for more details]{riley_thesis}. 

To monitor the time evolution of potentially varying parameters throughout each burst, we employ the procedure described in section 2.3 of \citet{kini:2023b}. This means that we first divide each burst into 8-time segments and then perform joint modelling of these segments. For this work, we assume that oscillations in each time segment originate from a single circular, uniformly emitting hot spot, as in \citet{kini:2023b}. Each hot spot is characterized by its temperature ($T_\mathrm{spotX}$, with X=1, 2, .., 8  the segment's number.), angular radius (or spot size; $\zeta_\mathrm{spotX}$), co-latitude ($\Theta_\mathrm{spotX}$) and phase shift ($\phi_\mathrm{spotX}$). For every burst, we treat $T_\mathrm{spot}$ and $\zeta_\mathrm{spot}$ as free parameters for each segment, without any predetermined order (unlike in \citealt{kini:2023b}). We hold the co-latitude and phase shift fixed across all segments of a burst (i.e. $\Theta_\mathrm{spot1}=\Theta_\mathrm{spot2}= ...=\Theta_\mathrm{spot8}=\Theta_\mathrm{spot}$, and the same for $\phi_\mathrm{spotX}$), but they are expected and allowed to vary from burst to burst. Additionally, we allow the rest of the star to emit with a uniform temperature $T_\mathrm{star}$. Although $T_\mathrm{star}$ is free to vary between each segment, we impose the trivial constraint ($T_\mathrm{spotX}$ > $T_\mathrm{starX}$) to expedite the sampling process.

\begin{figure}
    \centering
    \includegraphics[width=1.\columnwidth]{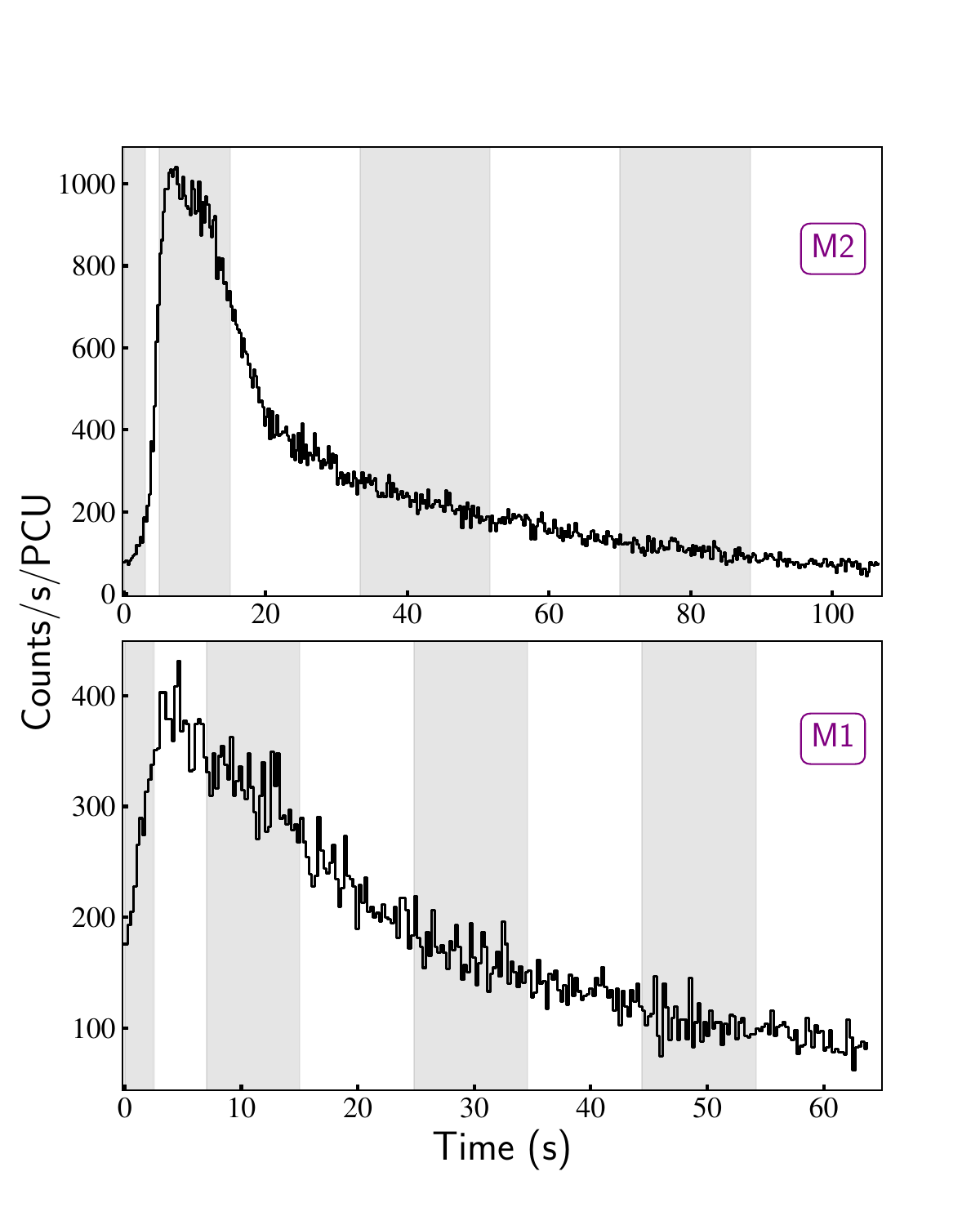}
      \caption{Illustrations of M1 and M2 bursts. The M1 burst corresponds to burst 6 of \J,\ while the M2 burst corresponds to burst 2. These light curves are shown with a time resolution of 0.25 seconds, covering the energy range from about 2 keV to about 26 keV.} The vertical bands denote each individual data segment.
    \label{fig:light_cures}
\end{figure}

In \citet{kini:2023b} each burst's rise was systematically divided into two segments of 2-seconds each, due to their consistent morphology. In this paper, we take a different approach. The rise is treated as either a single segment or divided in two, with varying time intervals across bursts. This distinction is due to the varying morphology of \J's\ bursts. Specifically, certain bursts exhibit either a low count rate or a rapid rise, which results in a low number of photons per segment when attempting to partition the burst's onset into two segments. We label the burst morphology as M1 when the rise is treated as one segment and as M2 when the rise is divided into two segments (see Table \ref{tab:data_table}). Due to the uniqueness of each burst, the duration of each segment is not standardized across all bursts, regardless of whether they are M1 bursts or M2 bursts. Each burst is segmented in a manner that ensures minimal variability in flux\footnote{"Minimal variability in flux" means that the changes in the light curve of the burst are not considered significant based on a visual inspection. This qualitative method was employed in \citep{kini:2023b} and led to accurate mass and radius inference. For further details, refer to footnote 8 of \citep{kini:2023b}} within each segment. In Figure \ref{fig:light_cures}, we present an example of the segmentation process for two bursts exhibiting distinct morphologies. Details on how the remaining bursts are segmented can be found on Zenodo (\url{{https://doi.org/10.5281/zenodo.8365643}}). The duration of each time segment is summarised in Table \ref{tab:duration} of Apppendix \ref{sec:appendix}.

\begin{table}
\centering

\begin{minipage}{1.\columnwidth}
\renewcommand*{\thempfootnote}{\fnsymbol{mpfootnote}}

\resizebox{\columnwidth}{!}{%
\begin{tabular}{p{0.9cm}cccccc}

\hline \hline
\multirow{2}{*}{Burst}   &OBSID &Start (days &  MINBAR ID\footnote{See \citet{Galloway:2020} or \url{https://burst.sci.monash.edu/}} & Duration & Count\footnote{The Counts refers to the number of X-ray event recorded for the time duration specified in the `Duration' column and the energy range outlined in Section \ref{sec:data}.} & Morphology   \\
\multicolumn{1}{l}{}    &       & since MJD 52790) &  & (s) &   \\

\hline

1 & 80138-04-02-00  &6.0468 & 3084 & 66.60  & 42736 &  M2 \\

 2 & 80138-04-03-00  & 7.2749& 3085  & 106.75 & 84106 & M2   \\

 3 & 80145-02-01-00  & 7.8899& 3086 & 101.00 & 82825 & M1  \\

 4 & 80418-01-01-01  &  9.2025& 3087 & 126.99 & 124670 &  M2   \\

 5 & 80418-01-01-03  &  10.1056& 3088 & 90.60  & 54317 & M2  \\

 6 & 80418-01-01-05  &  11.0356& 3089 & 63.99  & 35642&  M1\\ 

 7 & 80418-01-01-08  &  12.4727& 3091 & 106.75 & 113816 &  M2  \\

 8 & 80418-01-01-09  &  12.5697& 3092  & 75.24   & 44157 & M1  \\

 9 & 80418-01-02-01  &  13.0658& 3093  & 68.99 & 58716 & M2  \\

10 & 80418-01-02-00  &  13.7406& 3094 & 89.74  & 108361 & M2 \\

11 & 80418-01-02-03  &  14.0209& 3095 & 102.50  & 109697 & M2  \\

12 & 80418-01-02-06  &  15.7959& 3096  & 109.99  & 115745 & M2  \\

13 & 80418-01-02-07  &  16.7539& 3098 & 108.74 & 87581 &  M2 \\

14& 80418-01-02-07   &  16.8245& 3099 & 44.48   & 23673 & M1 \\

15& 80418-01-02-09   &  17.6734& 3100 & 86.74   & 110211 & M2  \\ 

16 & 80418-01-02-04  &  18.8018& 3101 & 95.24   & 87928 & M2  \\ 

17 & 80418-01-02-05  &  19.7881& 3102 & 88.50   & 86978 & M2  \\ 

18 & 80418-01-03-12 &  20.0747& 3103  & 94.00   & 111217 & M1 \\

19 & 80418-01-03-00 &  20.9079& 3104  & 49.73   & 29630 & M2  \\

20 & 80418-01-03-02  & 21.6460& 3105  & 111.24  & 90854 & M2   \\
21 & 80418-01-03-06  & 22.8951& 3106  & 172.77  & 101915 & M2 \\

22 & 80418-01-03-07  &  23.4752& 3107 & 129.60  & 125149 &  M2   \\

23 & 80418-01-04-00  &  27.7001& 3108 & 79.28   & 99358 & M2  \\

24 & 80418-01-04-00 &  27.8898 & 3109 & 105.49  & 90880 &  M2  \\

25 & 80418-01-04-01 &  28.8602 & 3110 & 90.56   & 54317 & M2  \\

26 & 80418-01-05-03 &  37.2599 & 3117 & 119.74  & 125175 & M2  \\

27 & 80418-01-05-08 &  38.7997& 3118  & 135.73  & 97346 &  M2  \\

28 & 80418-01-06-07 &  - & 3121   & - & -  &-  \\

\hline

\end{tabular}%
}
\caption{Burst properties overview. Burst 28 (which corresponds to Burst 27 in \citet[][]{Strohmayer:2003}) is excluded from this analysis because it shows evidence (although only marginal) of photospheric radius expansion which is not accounted for in our model \citep[see][]{Strohmayer:2003}.}
\label{tab:data_table}
\end{minipage}
\end{table}

In summary, we employ a single hot spot model and segment each burst into 8 time segments to capture the temporal changes of certain parameters.  As a result, for a given burst, the final model encompasses 31 free parameters in the sampling space:
\begin{itemize}
    \item Gravitational mass: $M$
    \item Equatorial radius: $R_\mathrm{eq}$
    \item Source-to-Earth distance: $D$
    \item Cosine of the observer's inclination: $\cos(i)$
    \item Hot spot  co-latitude: $\Theta_\mathrm{spot}$
    \item Hot spot phase : $\phi_\mathrm{spot}$
    \item Interstellar attenuation column density: $N_H$
    \item Hot spot temperature (8 in total): $T_\mathrm{spot1}, T_\mathrm{spot2}, ..., T_\mathrm{spot8}$
    \item Hot spot angular radius (8 in total): $\zeta_\mathrm{spot1}, \zeta_\mathrm{spot2}, ..., \zeta_\mathrm{spot8}$
    \item Star temperature (8 in total): $T_\mathrm{star1}, T_\mathrm{star2}, ..., T_\mathrm{star8}$
\end{itemize}

\subsubsection{Background assumptions}\label{sec:bkg}

By {\it background counts} we denote counts unrelated to astronomical sources originating solely from instrumental noise (instrumental background), as well as counts from other sources in the field of view, along with any additional sources of unpulsed emission from the X-ray binary itself. In PPM, the background model can significantly influence the inferred properties of the NS \citep[see e.g.][ for an in-depth discussion]{Salmi:2022}. Since the background model and counts during a burst are not well constrained, in this analysis, we employed two approaches.

In the \texttt{Bkg free} case, we refrained from imposing a specific model for the background, allowing it to vary freely as a parameter for each RXTE instrument channel. This means the background count can range between 0 and the observed data. In the \texttt{Bkg constrained} case, we assumed that during the burst, the upper limit of the background exhibits the same spectrum as the pre-burst emission. Recognizing that bursts are expected to disrupt the accretion flow, resulting in temporal variations in background counts throughout the burst \citep{Worpel:2015},  we set the lower limit of the background to 0 and the upper limit to the pre-burst counts times a scaling factor, \texttt{fa}, which represents the persistent model normalization as defined in \citet{Worpel:2015}. This means the background count, in this case, can range between 0 and \texttt{fa} times the pre-burst count. To compute \texttt{fa} for each time segment, we used the average upper limit obtained from a previous \J\ spectrum analysis \citep{Worpel:2015}. Most \texttt{fa} values mostly lie between 2 and 5, with a maximum value of about 8.7 (segment 3 of burst 26), and lowest values close to 0.01 (segment 7 of burst 23). For the pre-burst level, we took the first 10s before the start of each burst.

\subsubsection{Higher-order images}\label{sec:ho_images}
For sufficiently compact stars, photons originating from the same point on the stellar surface can reach the observer via multiple trajectories. Thus, higher-order images are expected for $R_{\mathrm{pole}}c^2/GM < 3.52$ \citep[see][and references therein]{Bogdanov:2021} where $R_{\mathrm{pole}}$ is the polar radius. During the inference runs in the \texttt{Bkg free} case, we imposed no restrictions on the image order. Images are summed over until higher-order images are no longer visible. This is computationally expensive, particularly when exploring regions of high compactness. Hence, in the \texttt{Bkg constrained} case (for which computational cost turned out to be much higher), we conducted the inference runs without imposing any restrictions on the image order only for the first ten bursts. Due to computational limitations, we set the image order to 1 for the remaining bursts. This implies that only the primary image is computed. For Burst 1 we ran both and confirmed that the choice did not affect the results. 

\subsubsection{Posterior computation}\label{sec:post_comb}
Given the model (which incorporates the relativistic ray-tracing, ISM, atmosphere, hot spot properties, and instrument response), we explore the parameter space and gauge how probable parameters ($\theta = (M, R_{\rm eq}, D, \cos(i),  \phi_\mathrm{spot}, \Theta_\mathrm{spot},\zeta_\mathrm{spot1},...,\zeta_\mathrm{spot8}, T_\mathrm{spot1},..,T_\mathrm{spot8},\\  T_\mathrm{star1},...,T_\mathrm{star8}, N_H$) are, based on both prior beliefs (see Section \ref{sec:priors}) and the new evidence provided by the burst data. We employ Bayes' formalism:

\begin{equation}\label{eq1}
  p(\theta_i|d_i)=\frac{p(d_i|\theta_i)p(\theta_i)}{p(d_i)},  
\end{equation}
where $\theta_i$ is a model parameter vector for the $i^{\mathrm{th}}$ burst, $d_i$ the corresponding data. $p(\theta_i|d_i)$ the posterior probability of the parameters $\theta_i$ given the data $d_i$, $p(d_i|\theta_i)$ the likelihood of observing the data $d_i$ given the
parameters $\theta_i$, $p(\theta_i)$ the prior probability (or initial belief about $\theta_i$), and $p(d_i)$ the evidence. 

Ideally, conducting a comprehensive joint inference run involving all the bursts, while keeping parameters such as $M, R_\mathrm{eq}, D, \cos(i), N_H$ fixed across the bursts, with the remaining parameters allowed to vary within and across bursts, would be preferred. However, with 27 bursts for \J ,\ this would entail exploring a parameter space with 707 dimensions, which is computationally challenging using current Nested Sampling techniques. In practice, we compute the posterior probability of the $N$ bursts, using \texttt{X-PSI}\footnote{\url{https://github.com/xpsi-group/xpsi}} version v2.0.0. Within \texttt{X-PSI}, this process is carried out using \texttt{MultiNest} \citep{MultiNest_2008,MultiNest_2009,MultiNest_2019} coupled with \texttt{PyMultinest} \citep{pymultinest:2014}. Although certain parameters (like mass) will be the same for all bursts, we compute the posterior probability of each sampled parameter vector for each burst independently. We do this for the sake of computational efficiency. The resulting posteriors from the $N$ bursts are then combined to derive the main results presented in this paper.

To combine these posteriors, we follow the methodology outlined in section 2.4 of \citet{kini:2023b}. We compute the combined posterior probability of a common set of parameters $\alpha$ using  Equation 5 of \citet{kini:2023b}:
\begin{equation}\label{eq4}
%\begin{split}
   p(\alpha|\mathcal{D})  \propto \prod\limits_{i=1}^{N} p(\alpha|d_i),
%\end{split}
\end{equation}
where $d_i$ is the data corresponding to the $i^{\mathrm{th}}$ burst,  $\mathcal{D}= \{d_i\}_{i=1}^{N}$ the set of data corresponding to the $N$  bursts, $p(\alpha|d_i)$ the posterior of the parameters $\alpha$ given the data $d_i$, and $p(\alpha|\mathcal{D})$ the combined posterior of the parameters $\alpha$ given the $N$  bursts.

To combine the posterior distributions using Equation \ref{eq4}, \citet{kini:2023b} employed a mesh featuring 400 points for each element of $\alpha$ (where $\alpha = (M, R_\mathrm{eq})$ \footnote{Some of the remaining parameters were assumed to be well constrained a priori.}). However, using this grid resolution for $\alpha = (M, R_\mathrm{eq}, D, \cos(i), N_H)$ would entail computing the posteriors for $400^5$ points, rendering it computationally impractical. Alternatively, one could set $\alpha = (M, R_\mathrm{eq})$ and marginalize over the remaining parameters. Yet, this approach carries the risk of overlooking correlations between $M$, $R$, and the marginalized parameters\footnote{Footnote 11 of \citet{kini:2023b} only applies if there are no correlations between parameters or if the correlations are the same.}.

Hence, instead of relying on the grid method, we initially approximate $p(\alpha|d_i)$ using Gaussian Kernel Density Estimation (KDE) from \texttt{scipy} \citep{scipy}.  Unlike in \citet{kini:2023b}, where the KDE bandwidth was fixed at a specific value (\texttt{bw=0.1}) to estimate $p(\alpha|d_i)$, here we adopt Scott's rule of thumb from \texttt{scipy} \citep{Scott:1992}. This rule dynamically calculates the appropriate bandwidth for the KDE for each of the $N$ bursts, considering both the sample size and the data variance.
 Then, we define a new likelihood function, $\mathcal{L} \propto \prod\limits_{i=1}^{N} p(\alpha|d_i)$, with $p(\alpha|d_i)$ representing the probability density function obtained using \texttt{scipy}. We sample the prior space of $\alpha$ using the new likelihood to derive the posterior distributions of $\alpha$ using \texttt{MultiNest} and \texttt{PyMultiNest}.

We also checked that when setting $\alpha = (M, R_\mathrm{eq}, D)$, both the grid method and the sampling method yielded similar results, as a check of our new sampling approach. However, the computational costs are drastically different: about $5\times 10^4$ core hours for the grid method and 60 core hours for the sampling method.

In Table \ref{tab:settings} of Appendix \ref{sec:appendix}, we present the sampler settings used to explore the prior space during inference for each burst. Additionally, we detail the settings employed when combining bursts. When combining bursts, we explored the joint $(M, R_\mathrm{eq}, D, \cos(i), N_H)$  parameter space using two distinct settings. Initially, we conducted explorations with $2\times 10^3$ live points, followed by another round with $10^5$ live points. Both settings yielded similar outcomes (see Figure \ref{fig:compre_LP} in Appendix \ref{sec:appendix}).

\subsubsection{Prior choice}\label{sec:priors}

For \J,\ prior knowledge of the parameters of interest is not as robust as that seen in some of the RMP sources examined using PPM \citep[see e.g.][]{Riley:2021, Salmi:2022}. Consequently, we opt for a broad prior distribution for the majority of these parameters: $M, R_{\rm eq}, \cos(i), \phi_\mathrm{spot}, \Theta_\mathrm{spot}, \zeta_\mathrm{spot1},...,\zeta_\mathrm{spot8}, T_\mathrm{spot1},.., \\ T_\mathrm{spot8}, T_\mathrm{star1},...,T_\mathrm{star8}, N_H$. These prior choices are summarized in Table \ref{tab:priors}. We also apply implicit prior criteria described in greater detail in section 2.5 of \citet[][]{kini:2023a} to expedite parameter sampling. These implicit prior criteria are summarized as follows:

\begin{itemize}
    \item We discard samples with $R_{\rm pole}/r_g(M) < 2.9$  to ensure compliance with the causality limit \citep[see e.g.][]{Gandolfi:2012}. $R_{\rm pole}$ is the polar radius and $r_g(M)=GM/c^2$. 
    \item Samples with $\log g \notin [13.7,14.9]$ are rejected to align with the range of surface gravities defined within the atmosphere table.
\end{itemize}

For distance, the situation is complicated.  There is no Gaia distance for this source \citep{Gaia_Collaboration}.  A study of the optical counterpart in quiescence, by \citet{Baglio:2012}, suggests a distance $\sim$ 11 kpc but this is also quite uncertain and is calculated assuming a NS mass of  1.4 \msol. Another estimate can be obtained from the marginal detection of PRE in the brightest burst from this source (Burst 28) \citep{Strohmayer:2003}.  If the peak luminosity of the brightest burst can indeed be equated with the Eddington limit then the distance is given by \citep[see ][]{Galloway:2008}:

\begin{equation}
\begin{split}
d =8.6 \left(\frac{F_\mathrm{{peak}}}{3 \times 10^{-8} \rm erg\,cm^{-2}\,s^{-1}} \right)^{-1/2} \left(\frac{M}{1.4 \mathrm{M_\odot}}\right)^{1/2}  \\ \times \left(\frac{1+z(M,R)}{1.31}\right)^{-1/2} \left(1+X \right)^{-1/2} \mathrm{kpc}
\end{split}
\end{equation}
where $F_\mathrm{{peak}}$ is the peak flux of the PRE burst,  $M$ the NS mass, $R$ ($R\approx R_{\mathrm{eq}}$) the NS radius, $1+z(M,R)=(1-2GM/Rc^2)^{-1/2}$ the  gravitational redshift, and $X$ the hydrogen mass fraction.

For a marginal detection of PRE, where the peak flux of the brightest burst might not reach the Eddington limit, this yields instead an upper limit on distance. Considering the bounds of the prior distribution for both mass and radius and under the assumption of the extreme scenario where burst 28 exclusively involves pure Helium burning (\texttt{X=0.0}), we have determined an estimated distance of approximately 14 kpc. Hence, we chose this as an upper limit for the distance. For the lower bounds for the distance, we opt for 3 kpc. This is motivated by the expectation that \J\ is likely to be located no closer than 3.8 kpc \citep{Krauss:2005sj}. This lower limit on the distance was obtained by fitting the phase-averaged X-ray spectrum of \J\ in outburst.

   \begin{table}
    \renewcommand*{\thempfootnote}{\fnsymbol{mpfootnote}}
       %\centering
       \caption{ Parameters and their respective prior density used for sampling.}
       \begin{minipage}{1.\columnwidth}
       %\centering
        \begin{tabular}{p{3.5cm} l}
        \hline \hline
          Parameter                                                        & Prior density \\ \hline
 
          $M$ ($M_{\odot}$)                                                & $M\sim\mathcal{U}(1.0,3.0)$                   \\ 
          $R_{\rm eq}$ (km)                                                &$R_{\rm eq}\sim\mathcal{U}(3r_g(1.0),16.0)$\footnote{$r_g(1.0)$: Solar Schwarzschild gravitational radius.} \\
          $D$ (kpc)                                                        & $D\sim\mathcal{U}(3.0,14.0)$ \\
          $\cos(i)$                                                        &$\cos(i)\sim\mathcal{U}(0.0,1.0)$                 \\
          $\phi_\mathrm{spot}$ (cycles)                                    &$\phi_\mathrm{spot}\sim\mathcal{U}(-0.25,0.75)$   \\
          $\Theta_\mathrm{spot}$ (radian)                                  &$\cos(\Theta_\mathrm{spot}) \sim\mathcal{U}(-1.0,1.0)$ \\
          $\zeta_\mathrm{spotX}$ (rad)                                  &$\zeta_\mathrm{spot}\sim\mathcal{U}(0.0,\pi/2)$    \\
          $\log[T_\mathrm{spotX}(\mathrm{K})/1\mathrm{K}]$                 &$\log[T_\mathrm{spot}(\mathrm{K})/1\mathrm{K}]\sim\mathcal{U}(6.7,7.6)$\footnote{Temperature bounds set by bursting atmosphere table computed as described in more detail in \citet[see][]{kini:2023a}.}  \\
          $\log[T_\mathrm{starX}(\mathrm{K})/1\mathrm{K}]$                 &$\log[T_\mathrm{star}(\mathrm{K})/1\mathrm{K}]\sim\mathcal{U}(6.7,7.6)$           \\
          $N_{H}$   ($10^{20}\mathrm{cm}^{-2}$)                            &$N_{H} \sim \mathcal{U}(9.0,100.0)$                  \\ \hline
       \end{tabular}
       \end{minipage}
       
       \label{tab:priors}
       \end{table}

%% file: Sec2-Results.tex
\section{Results}\label{sec:result}
We conducted a total of 54 inference runs, with 27 each for both the \texttt{Bkg free} and \texttt{Bkg constrained} cases. Of these, 53 runs were completed successfully. However, Burst 23 in the \texttt{Bkg constrained} case had to be halted due to excessive computational demand. Imposing a constraint on the background during inference mostly leads to a significantly higher number of likelihood evaluations before the run converges, resulting in longer run times. Therefore, the results for the combined bursts do not include Burst 23 for either of the cases. The total number of counts excluding Burst 23 is 2197642. The total computing time is about 2.4 million and 4.8 million core hours (excluding burst 23 run time) respectively for the \texttt{Bkg free} and \texttt{Bkg constrained} cases. The details of runtimes for each burst are highlighted in Table \ref{tab:core_hours} of Appendix \ref{sec:appendix}.

In this section, we first assess the one-hot-spot model quality in Section \ref{sec:model_q}. Starting from Section \ref{sec:common}, we highlight the properties of \J\, along with the characteristics of the bursts and oscillations that we obtain from PPM. We first cover the parameters known to be common to all bursts: mass, radius, distance, observer inclination, and column density. Then we present the findings for the parameters that vary with time during the burst and from burst to burst. 

\subsection{Model quality}\label{sec:model_q}

\begin{figure}%[h]
    \centering
    \includegraphics[width=1.\columnwidth]{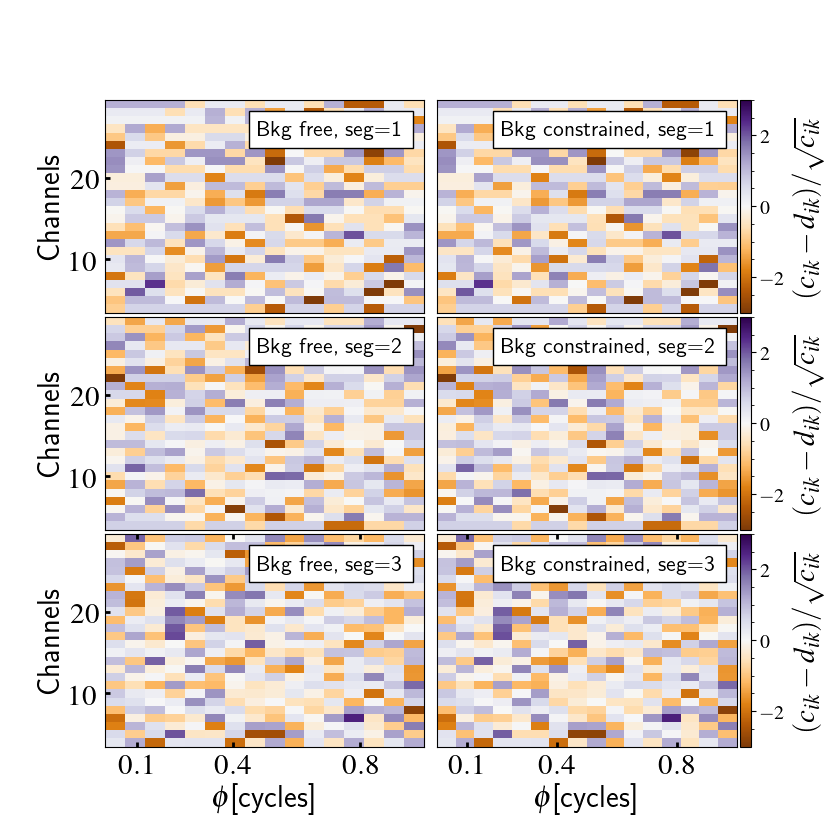}
      \caption{Residual plots for Burst 1 for the first three segments for \texttt{Bkg free} (left) and \texttt{Bkg constrained} (right). The residuals are the difference between the model counts (for the maximum likelihood solution) and the data counts, normalized by the model count counts in each instrument energy channel and phase bin.  $c_{ik}$ and $d_{ik}$ denote respectively the model counts and the data counts in the $i^{th}$ rotation phase and  $k^{th}$ energy channel.}
    \label{fig:residuals}
\end{figure}

\begin{figure}%[h]
    \centering
    \includegraphics[width=1.\columnwidth]{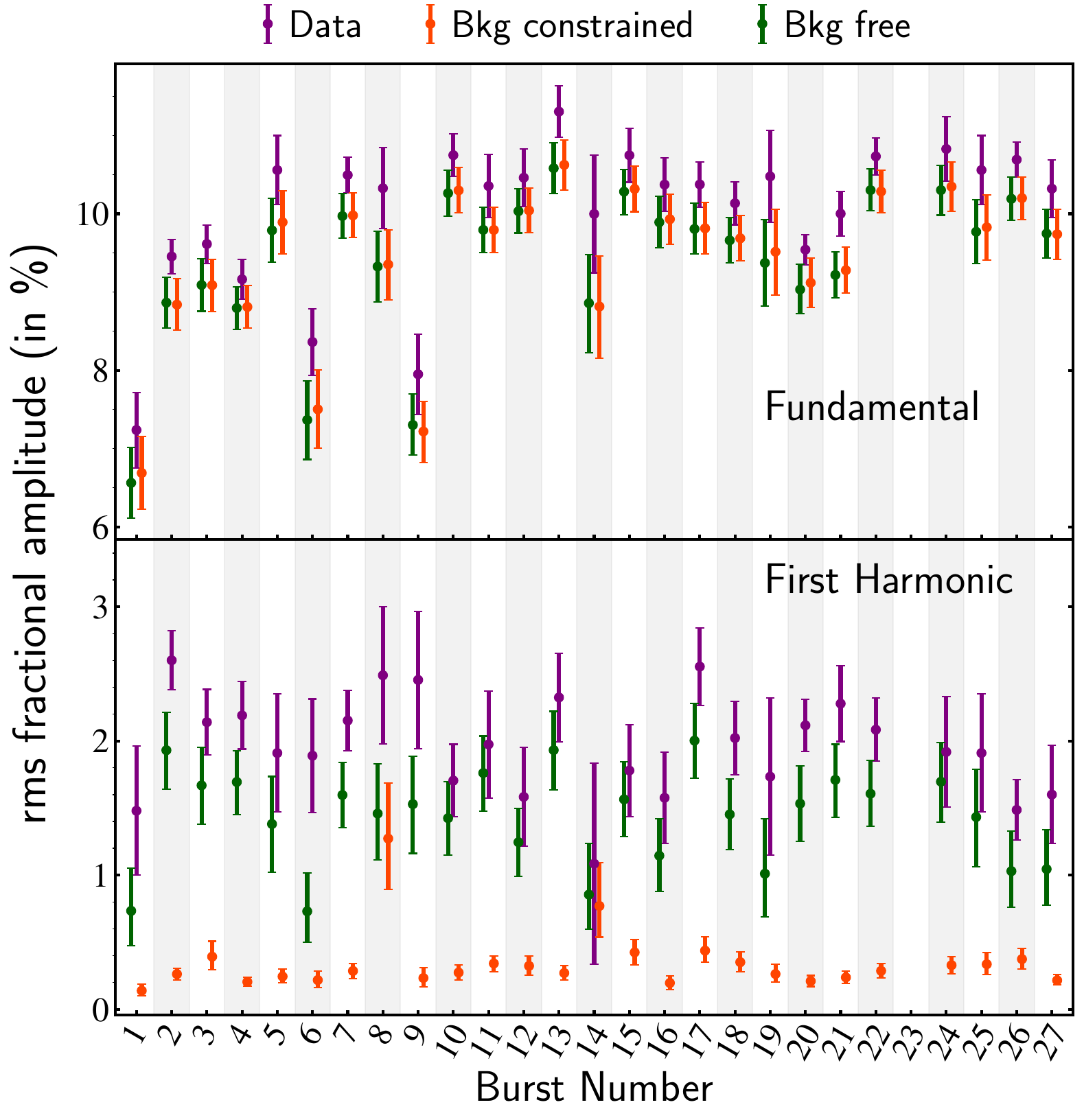}
      \caption{Medians and 68\% credible intervals of the probability distribution of the bolometric (combined across segments) rms FA of the single hot spot model for both the fundamental (top) and the first harmonic (bottom) in each burst in the \texttt{Bkg free} and \texttt{Bkg constrained} cases alongside rms FA in the data.}% all the posterior samples for each burst.}
    \label{fig:harmonics}
\end{figure}

In Figure \ref{fig:residuals}, we present the residual plots for Burst 1 for both the \texttt{Bkg free} and \texttt{Bkg constrained} cases for the first three segments. These residuals are computed using the maximum likelihood solution in each case. Residuals for the remaining segments and bursts are available on Zenodo \textcolor{purple}{(\url{{https://doi.org/10.5281/zenodo.8365643}})}. We also show the distribution of the residuals for Burst 1 in Figure \ref{fig:residuals_distibutions} of Appendix \ref{sec:appendix}. Notably, none of the residuals reveal discernible patterns or clustering, or a deviation from the overall expected distribution, either of which might suggest that the model's performance is inadequate. However, evaluating the model quality solely based on residuals may not provide a complete assessment.

Comparing the medians (and 68\% credible intervals) of the probability distribution of the bolometric (combined across segments) root mean square fractional amplitude (rms FA) from the (single hot spot) model for both the fundamental and the first harmonic to that of the data, as illustrated in Fig \ref{fig:harmonics}, reveals interesting insights. Across both the \texttt{Bkg free} and \texttt{Bkg constraints} cases, the rms FAs of the fundamental are almost all consistent, within 68\% credible intervals, with the rms FAs observed in the data. This suggests that the model can reproduce the rms FA of the fundamental in the data. However, it is notable the rms FAs in both \texttt{Bkg free} and \texttt{Bkg constrained} cases are consistently lower than those observed in the data. A similar trend is observable in the \texttt{Bkg free} case for the second harmonic for most bursts. However, for the first harmonic, in the \texttt{Bkg constraints} case, rms FAs are significantly lower than the values observed in the data. The rms FA of all the bursts, except from Burst 14, cannot reproduce the rms FA observed in the data. These suggest that the single hot spot model, particularly under the \texttt{Bkg constrained} scenario,  struggles to accurately model the data.

\subsection{Common parameters shared across all bursts}\label{sec:common}

In Table \ref{tab:each_post}, we show the inferred values of the mass, radius, distance, observer’s inclination, and column density for each burst for both the \texttt{Bkg free} and the \texttt{Bkg constrained} cases. The table also shows the combined result in the last row. The provided values are the median of the posterior distribution and the 68\% credible interval. These values are derived using the samples and their corresponding posteriors obtained through the method described in Section \ref{sec:post_comb}. The complete posterior plots for each burst can be accessed on Zenodo (\url{{https://doi.org/10.5281/zenodo.8365643}}). We found that, for the initial runs where we leave the background free (\texttt{Bkg free}),  the average uncertainties of the upper (lower) 68\% credible interval for the mass and radius, for a single burst are $+26\%$($-19\%$) and  $+18\%$($-16\%$) respectively. For the \texttt{Bkg constrained} case, the average uncertainties are  $+26\%$($-12\%$) and  $+22\%$($-22\%$) respectively for the mass and radius. 
The results also show that constraining the background systematically leads to smaller distances and inclinations compared to when the background is left unconstrained.

\begin{figure*}%[h]
    \centering
    \includegraphics[width=2.\columnwidth]{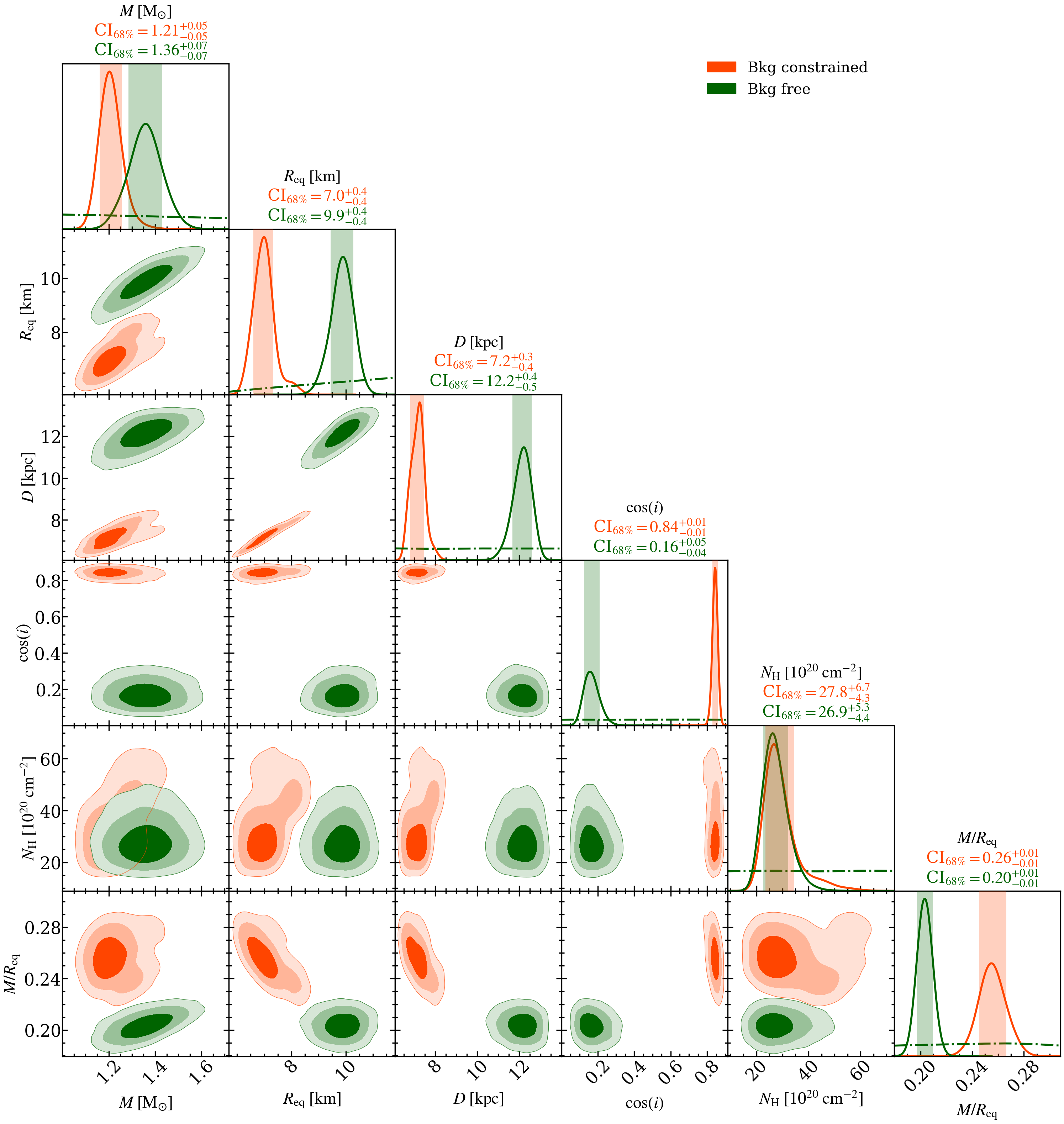}
      \caption{Combined posterior distributions of mass, radius, distance, observer inclination, column density, and compactness. The two-dimensional posterior distributions are, from most opaque to least, the 68\%, 95\%, and 99\% posterior credible region. The solid (dash-dotted) lines along the diagonal represent the marginalized posterior (prior) distribution of each parameter. The vertical bands are the inferred 68\% credible intervals.}
    \label{fig:combed_MR}
\end{figure*}

In Figure \ref{fig:combed_MR}, we present the combined posterior distribution plots of the common parameters, using information from the 26 bursts. The combined posteriors for each subset (M1 and M2) are shown in Figure \ref{fig:subset} of Appendix\ref{sec:appendix}. The two-dimensional posterior distributions are, from most opaque to least, the 68\%, 95\%, and 99\% posterior credible region. The solid (dash-dotted) lines in the 1D credible interval panels represent each parameter's marginalized posterior (prior) distribution. The vertical bands are the inferred 68\% credible intervals. 

 A prominent observation is the emergence of a multimodal likelihood surface. While 3 modes were identified using $2\times 10^3$ live points, this number increased to 9 when employing $10^5$ live points. However, irrespective of the sampler settings, the preferred solution remains consistent based on the evidence (Table \ref{tab:live_points} of the Appendix \ref{sec:appendix} for an overview of the Maximum A Posteriori (MAP) for each mode). In this section, we present the posterior distributions obtained using  $2\times 10^3$ live points.

For the scenario with unconstrained background, the inferred values for mass, radius, and distance are $1.36^{+0.07}_{-0.07}$ \msol, $9.9^{+0.4}_{-0.4}$ km, and $12.2^{+0.4}_{-0.5}$ kpc, respectively. The uncertainties (the full 68\% credible interval) associated with mass and radius are approximately $\Delta M/M \approx 10\% $ and $\Delta R_\mathrm{eq}/R_\mathrm{eq}= 8\%$, respectively. The corresponding values for $\cos(i)$ and $N_H$ are $0.16^{+0.05}_{-0.04}$ and $26.9^{+5.3}_{-4.4}\times 10^{20}\mathrm{cm}^{-2}$. The overall inferred background is very high, for more discussion see Section \ref{sec:discussin_initial_runs}.

Conversely, under the constraint of an upper limit on the background (\texttt{Bkg constrained}), the inferred parameters adjust to $1.21^{+0.05}_{-0.05}$ \msol for mass, $7.0^{+0.4}_{-0.4}$ km for radius, and $7.2^{+0.3}_{-0.4}$ kpc for distance. This time, the uncertainties in mass and radius are approximately $\Delta M/M \approx 8\% $ and $\Delta R_\mathrm{eq}/R_\mathrm{eq}= 11\%$. The inferred values for $\cos(i)$ and $N_H$ become $0.84^{+0.01}_{-0.01}$ and $27.84^{+6.72}_{-4.34}\times 10^{20}\mathrm{cm}^{-2}$, respectively.

While the combined estimations of the mass, and $N_H$ do not exhibit inconsistency between the \texttt{Bkg free} and \texttt{Bkg constrained} cases, notable disparities emerge in the inferred values for radius, distance, and $\cos(i)$. In the \texttt{Bkg constrained} case, we inferred a lower radius and distance and a smaller observer inclination.

\begin{table*}
\centering
%\resizebox{2\columnwidth}{!}{%
\begin{minipage}{1.\textwidth}
\renewcommand*{\thempfootnote}{\fnsymbol{mpfootnote}}
\begin{tabular}{cccccccccccc}

\hline \hline 

\multirow{2}{*}{Burst} & \multicolumn{2}{c}{$M$ ($M_{\odot}$)} & \multicolumn{2}{c}{$R_{\rm eq}$ (km)} & \multicolumn{2}{c}{$D$ (kpc)} & \multicolumn{2}{c}{$\cos(i)$}      & \multicolumn{2}{c}{$N_H$ ($10^{20}\mathrm{cm}^{-2}$)} \\

\cmidrule(lr){2-3}
\cmidrule(lr){4-5}
\cmidrule(lr){6-7}
\cmidrule(lr){8-9}
\cmidrule(lr){10-11}

 \multicolumn{1}{c}{} & 
%  \multicolumn{1}{c}{\textcolor{darkgreen}{$\star$}} &  
%  \multicolumn{1}{c}{\textcolor{orange}{$\dag$}} &
%  \multicolumn{1}{c}{\textcolor{darkgreen}{$\star$}} &
%  \multicolumn{1}{c}{\textcolor{orange}{$\dag$}} &
%  \multicolumn{1}{c}{\textcolor{darkgreen}{$\star$}} &
%  \multicolumn{1}{c}{\textcolor{orange}{$\dag$}} &
%  \multicolumn{1}{c}{\textcolor{darkgreen}{$\star$}} &
%  \multicolumn{1}{c}{\textcolor{orange}{$\dag$}} &
%  \multicolumn{1}{c}{\textcolor{darkgreen}{$\star$}} &
%  \multicolumn{1}{c}{\textcolor{orange}{$\dag$}} &

\multicolumn{1}{c}{ \texttt{Bkg free}} & 
 \multicolumn{1}{c}{ \texttt{Bkg con}} &  
  \multicolumn{1}{c}{ \texttt{Bkg free}} & 
 \multicolumn{1}{c}{ \texttt{Bkg con}} &  
  \multicolumn{1}{c}{ \texttt{Bkg free}} & 
 \multicolumn{1}{c}{ \texttt{Bkg con}} &  
  \multicolumn{1}{c}{ \texttt{Bkg free}} & 
 \multicolumn{1}{c}{ \texttt{Bkg con}} &  
  \multicolumn{1}{c}{ \texttt{Bkg free}} & 
 \multicolumn{1}{c}{ \texttt{Bkg con}} &  
  
   \\

\hline
\vr

 1 & \multicolumn{1}{c}{$1.52^{+ 0.47}_{-0.33}$} &  \multicolumn{1}{c}{$1.57^{+ 0.48}_{-0.36}$}  &  \multicolumn{1}{c}{$10.06^{+ 2.23}_{-2.03}$}  & \multicolumn{1}{c}{$11.42^{+ 2.55}_{-2.70}$}  & \multicolumn{1}{c}{$10.95^{+ 1.89}_{-2.16}$}  & \multicolumn{1}{c}{$9.85^{+ 2.00}_{-2.05}$}  &  \multicolumn{1}{c}{$0.35^{+0.26}_{-0.22}$} & \multicolumn{1}{c}{$0.86^{+0.07}_{-0.13}$} &   \multicolumn{1}{c}{$50.51^{+29.19}_{-26.23}$} &  \multicolumn{1}{c}{$44.56^{+30.11}_{-23.04}$} \\ \vr

 2 & \multicolumn{1}{c}{$1.36^{+ 0.36}_{-0.23}$} &  \multicolumn{1}{c}{$1.73^{+ 0.46}_{-0.39}$}  &  \multicolumn{1}{c}{$10.93^{+ 1.77}_{-1.57}$}  & \multicolumn{1}{c}{$11.01^{+ 2.32}_{-2.58}$}  & \multicolumn{1}{c}{$12.41^{+ 1.07}_{-1.61}$}  & \multicolumn{1}{c}{$10.65^{+1.80}_{-2.05}$}  &  \multicolumn{1}{c}{$0.26^{+0.21}_{-0.16}$} & \multicolumn{1}{c}{$0.83^{+0.08}_{-0.13}$} &   \multicolumn{1}{c}{$45.26^{+31.25}_{-23.95}$} &  \multicolumn{1}{c}{$44.81^{+30.11}_{-23.26}$} \\ \vr

 3 & \multicolumn{1}{c}{$1.33^{+ 0.34}_{-0.22}$} &  \multicolumn{1}{c}{$1.37^{+ 0.39}_{-0.24}$}  &  \multicolumn{1}{c}{$10.69^{+ 1.83}_{-1.65}$}  & \multicolumn{1}{c}{$9.39^{+ 2.73}_{-2.23}$}  & \multicolumn{1}{c}{$11.92^{+ 1.38}_{-1.93}$}  & \multicolumn{1}{c}{$9.88^{+ 2.06}_{-1.75}$}  &  \multicolumn{1}{c}{$0.28^{+0.23}_{-0.18}$} & \multicolumn{1}{c}{$0.74^{+0.13}_{-0.23}$} &   \multicolumn{1}{c}{$42.75^{+31.93}_{-22.81}$} &  \multicolumn{1}{c}{$36.14^{+29.65}_{-18.02}$} \\ \vr

 4 & \multicolumn{1}{c}{$1.65^{+ 0.40}_{-0.35}$} &  \multicolumn{1}{c}{$1.83^{+ 0.37}_{-0.39}$}  &  \multicolumn{1}{c}{$12.06^{+ 1.74}_{-1.74}$}  & \multicolumn{1}{c}{$13.54^{+ 1.57}_{-2.23}$}  & \multicolumn{1}{c}{$12.49^{+ 1.03}_{-1.59}$}  & \multicolumn{1}{c}{$10.42^{+ 1.03}_{-1.47}$}  &  \multicolumn{1}{c}{$0.24^{+0.19}_{-0.15}$} & \multicolumn{1}{c}{$0.86^{+0.05}_{-0.08}$} &   \multicolumn{1}{c}{$47.77^{+31.25}_{-25.54}$} &  \multicolumn{1}{c}{$54.39^{+28.05}_{-28.28}$} \\ \vr

 5 & \multicolumn{1}{c}{$1.40^{+ 0.34}_{-0.25}$} &  \multicolumn{1}{c}{$1.40^{+ 0.31}_{-0.24}$}  &  \multicolumn{1}{c}{$9.48^{+ 1.68}_{-1.48}$}  & \multicolumn{1}{c}{$8.95^{+ 1.88}_{-1.97}$}  & \multicolumn{1}{c}{$12.37^{+ 1.10}_{-1.66}$}  & \multicolumn{1}{c}{$11.01^{+1.79}_{-1.94}$}  &  \multicolumn{1}{c}{$0.28^{+0.23}_{-0.18}$} & \multicolumn{1}{c}{$0.77^{+0.10}_{-0.16}$} &   \multicolumn{1}{c}{$51.19^{+30.11}_{-27.37}$} &  \multicolumn{1}{c}{$63.51^{+23.95}_{-31.47}$} \\ \vr

 6 & \multicolumn{1}{c}{$1.76^{+ 0.50}_{-0.42}$} &  \multicolumn{1}{c}{$1.74^{+ 0.54}_{-0.43}$}  &  \multicolumn{1}{c}{$10.29^{+ 2.41}_{-2.12}$}  & \multicolumn{1}{c}{$9.42^{+ 2.46}_{-2.35}$}  & \multicolumn{1}{c}{$11.83^{+ 1.44}_{-2.04}$}  & \multicolumn{1}{c}{$10.25^{+2.22}_{-2.34}$}  &  \multicolumn{1}{c}{$0.39^{+0.26}_{-0.24}$} & \multicolumn{1}{c}{$0.76^{+0.14}_{-0.25}$} &   \multicolumn{1}{c}{$52.33^{+29.65}_{-27.60}$} &  \multicolumn{1}{c}{$50.28^{+30.56}_{-26.46}$} \\ \vr

 7 & \multicolumn{1}{c}{$1.70^{+ 0.37}_{-0.35}$} &  \multicolumn{1}{c}{$1.66^{+0.45}_{-0.38}$}  &  \multicolumn{1}{c}{$11.45^{+ 1.71}_{-1.77}$}  & \multicolumn{1}{c}{$11.59^{+ 2.55}_{-2.73}$}  & \multicolumn{1}{c}{$12.07^{+ 1.30}_{-1.83}$}  & \multicolumn{1}{c}{$9.34^{+1.73}_{-1.90}$}  &  \multicolumn{1}{c}{$0.25^{+0.21}_{-0.16}$} & \multicolumn{1}{c}{$0.83^{+0.08}_{-0.13}$} &   \multicolumn{1}{c}{$42.98^{+31.70}_{-22.81}$} &  \multicolumn{1}{c}{$44.81^{+31.47}_{-23.72}$} \\ \vr

 8 & \multicolumn{1}{c}{$1.42^{+ 0.41}_{-0.27}$} &  \multicolumn{1}{c}{$1.41^{+ 0.38}_{-0.26}$}  &  \multicolumn{1}{c}{$10.03^{+2.06}_{-1.77}$}  & \multicolumn{1}{c}{$10.06^{+2.09}_{-1.86}$}  & \multicolumn{1}{c}{$11.74^{+ 1.48}_{-1.99}$}  & \multicolumn{1}{c}{$11.72^{+ 1.48}_{-1.94}$}  &  \multicolumn{1}{c}{$0.30^{+0.25}_{-0.19}$} & \multicolumn{1}{c}{$0.32^{+0.25}_{-0.21}$} &   \multicolumn{1}{c}{$46.40^{+31.25}_{-24.86}$} &  \multicolumn{1}{c}{$46.40^{+31.25}_{-24.86}$} \\ \vr

 9 & \multicolumn{1}{c}{$1.49^{+ 0.57}_{-0.33}$} &  \multicolumn{1}{c}{$1.85^{+ 0.37}_{-0.41}$}  &  \multicolumn{1}{c}{$12.20^{+ 2.03}_{-2.06}$}  & \multicolumn{1}{c}{$11.56^{+ 1.97}_{-2.73}$}  & \multicolumn{1}{c}{$12.20^{+ 1.20}_{-1.84}$}  & \multicolumn{1}{c}{$11.86^{+1.43}_{-2.29}$}  &  \multicolumn{1}{c}{$0.26^{+0.22}_{-0.17}$} & \multicolumn{1}{c}{$0.77^{+0.11}_{-0.16}$} &   \multicolumn{1}{c}{$49.82^{+30.33}_{-26.46}$} &  \multicolumn{1}{c}{$49.37^{+30.33}_{-26.00}$} \\ \vr

10 & \multicolumn{1}{c}{$1.48^{+ 0.36}_{-0.29}$} &  \multicolumn{1}{c}{$1.41^{+ 0.40}_{-0.26}$}  &  \multicolumn{1}{c}{$11.07^{+ 1.83}_{-1.94}$}  & \multicolumn{1}{c}{$10.90^{+2.61}_{-2.44}$}  & \multicolumn{1}{c}{$11.49^{+ 1.59}_{-1.97}$}  & \multicolumn{1}{c}{$8.84^{+1.74}_{-1.63}$}  &  \multicolumn{1}{c}{$0.30^{+0.23}_{-0.19}$} & \multicolumn{1}{c}{$0.82^{+0.09}_{-0.15}$} &   \multicolumn{1}{c}{$43.67^{+31.93}_{-23.26}$} &  \multicolumn{1}{c}{$41.84^{+31.25}_{-21.89}$} \\ \vr

11 & \multicolumn{1}{c}{$1.38^{+ 0.33}_{-0.24}$} &  \multicolumn{1}{c}{$1.96^{+0.45}_{-0.45}$}  &  \multicolumn{1}{c}{$11.30^{+ 1.74}_{-1.59}$}  & \multicolumn{1}{c}{$13.80^{+1.45}_{-2.29}$}  & \multicolumn{1}{c}{$12.36^{+ 1.10}_{-1.67}$}  & \multicolumn{1}{c}{$11.29^{+1.22}_{-1.57}$}  &  \multicolumn{1}{c}{$0.27^{+0.23}_{-0.18}$} & \multicolumn{1}{c}{$0.81^{+0.08}_{-0.11}$} &   \multicolumn{1}{c}{$41.16^{+31.70}_{-21.89}$} &  \multicolumn{1}{c}{$39.79^{+31.47}_{-20.98}$} \\ \vr

12 & \multicolumn{1}{c}{$1.24^{+ 0.26}_{-0.16}$} &  \multicolumn{1}{c}{$1.29^{+0.33}_{-0.19}$}  &  \multicolumn{1}{c}{$9.45^{+ 1.74}_{-1.59}$}  & \multicolumn{1}{c}{$9.39^{+ 2.78}_{-2.12}$}  & \multicolumn{1}{c}{$11.00^{+ 1.83}_{-2.03}$}  & \multicolumn{1}{c}{$8.56^{+2.03}_{-1.72}$}  &  \multicolumn{1}{c}{$0.33^{+0.25}_{-0.21}$} & \multicolumn{1}{c}{$0.79^{+0.09}_{-0.16}$} &   \multicolumn{1}{c}{$40.70^{+31.70}_{-21.44}$} &  \multicolumn{1}{c}{$39.11^{+30.79}_{-20.30}$} \\ \vr

13 & \multicolumn{1}{c}{$1.36^{+ 0.32}_{-0.23}$} &  \multicolumn{1}{c}{$1.23^{+0.26}_{-0.15}$}  &  \multicolumn{1}{c}{$10.75^{+ 1.74}_{-1.65}$}  & \multicolumn{1}{c}{$8.81^{+2.12}_{-1.74}$}  & \multicolumn{1}{c}{$11.72^{+ 1.47}_{-1.91}$}  & \multicolumn{1}{c}{$9.16^{+1.74}_{-1.54}$}  &  \multicolumn{1}{c}{$0.25^{+0.21}_{-0.16}$} & \multicolumn{1}{c}{{$0.80^{+0.10}_{-0.17}$}} &   \multicolumn{1}{c}{$37.51^{+30.79}_{-19.39}$} &  \multicolumn{1}{c}{{$31.81^{+26.68}_{-15.51}$}} \\ \vr

14& \multicolumn{1}{c}{$1.52^{+ 0.45}_{-0.32}$} &  \multicolumn{1}{c}{$1.56^{+0.45}_{-0.33}$}  &  \multicolumn{1}{c}{$8.58^{+ 2.17}_{-1.80}$}  & \multicolumn{1}{c}{$8.37^{+2.17}_{-1.80}$}  & \multicolumn{1}{c}{$11.14^{+ 1.85}_{-2.56}$}  & \multicolumn{1}{c}{$10.96^{+1.95}_{-2.57}$}  &  \multicolumn{1}{c}{$0.32^{+0.26}_{-0.21}$} & \multicolumn{1}{c}{{$0.34^{+0.27}_{-0.22}$}} &   \multicolumn{1}{c}{$49.60^{+29.88}_{-26.23}$} &  \multicolumn{1}{c}{{$49.82^{+30.11}_{-26.46}$}} \\ \vr

15& \multicolumn{1}{c}{$1.44^{+ 0.37}_{-0.28}$} &  \multicolumn{1}{c}{$1.59^{+0.47}_{-0.36}$}  &  \multicolumn{1}{c}{$11.30^{+ 1.91}_{-1.86}$}  & \multicolumn{1}{c}{$12.26^{+2.26}_{-2.81}$}  & \multicolumn{1}{c}{$11.86^{+ 1.41}_{-1.86}$}  & \multicolumn{1}{c}{$10.41^{+1.69}_{-2.05}$}  &  \multicolumn{1}{c}{$0.31^{+0.24}_{-0.20}$} & \multicolumn{1}{c}{$0.78^{+0.10}_{-0.15}$} &   \multicolumn{1}{c}{$45.04^{+31.70}_{-23.95}$} &  \multicolumn{1}{c}{$42.98^{+31.70}_{-22.81}$} \\ \vr

16 & \multicolumn{1}{c}{$1.23^{+ 0.24}_{-0.16}$} &  \multicolumn{1}{c}{$1.26^{+0.30}_{-0.17}$}  &  \multicolumn{1}{c}{$9.13^{+ 1.45}_{-1.33}$}  & \multicolumn{1}{c}{$8.81^{+2.32}_{-1.83}$}  & \multicolumn{1}{c}{$11.97^{+ 1.34}_{-1.79}$}  & \multicolumn{1}{c}{$8.83^{+1.84}_{-1.51}$}  &  \multicolumn{1}{c}{$0.30^{+0.24}_{-0.19}$} & \multicolumn{1}{c}{$0.82^{+0.09}_{-0.15}$} &   \multicolumn{1}{c}{$47.54^{+31.02}_{-25.32}$} &  \multicolumn{1}{c}{$46.18^{+30.56}_{-24.40}$} \\ \vr

17 & \multicolumn{1}{c}{$1.30^{+ 0.30}_{-0.20}$} &  \multicolumn{1}{c}{$1.35^{+0.35}_{-0.22}$}  &  \multicolumn{1}{c}{$11.10^{+ 2.00}_{-1.74}$}  & \multicolumn{1}{c}{$8.72^{+3.31}_{-2.46}$}  & \multicolumn{1}{c}{$11.47^{+ 1.59}_{-1.91}$}  & \multicolumn{1}{c}{$8.69^{+2.17}_{-1.78}$}  &  \multicolumn{1}{c}{$0.32^{+0.22}_{-0.20}$} & \multicolumn{1}{c}{$0.75^{+0.17}_{-0.36}$} &   \multicolumn{1}{c}{$49.82^{+30.33}_{-26.46}$} &  \multicolumn{1}{c}{$40.93^{+31.02}_{-21.21}$} \\ \vr

18 & \multicolumn{1}{c}{$1.41^{+ 0.34}_{-0.25}$} &  \multicolumn{1}{c}{$1.47^{+0.41}_{-0.29}$}  &  \multicolumn{1}{c}{$10.49^{+ 1.86}_{-1.77}$}  & \multicolumn{1}{c}{$12.61^{+2.15}_{-2.49}$}  & \multicolumn{1}{c}{$11.37^{+ 1.66}_{-1.91}$}  & \multicolumn{1}{c}{$10.03^{+1.50}_{-1.59}$}  &  \multicolumn{1}{c}{$0.26^{+0.21}_{-0.17}$} & \multicolumn{1}{c}{$0.81^{+0.07}_{-0.10}$} &   \multicolumn{1}{c}{$38.42^{+31.47}_{-19.84}$} &  \multicolumn{1}{c}{$37.51^{+31.02}_{-19.39}$} \\ \vr

19 & \multicolumn{1}{c}{$1.58^{+ 0.47}_{-0.35}$} &  \multicolumn{1}{c}{$1.71^{+0.40}_{-0.37}$}  &  \multicolumn{1}{c}{$9.21^{+ 2.15}_{-1.80}$}  & \multicolumn{1}{c}{$9.01^{+1.89}_{-2.03}$}  & \multicolumn{1}{c}{$12.03^{+ 1.33}_{-2.00}$}  & \multicolumn{1}{c}{$11.10^{+1.84}_{-2.25}$}  &  \multicolumn{1}{c}{$0.33^{+0.25}_{-0.21}$} & \multicolumn{1}{c}{$0.74^{+0.13}_{-0.22}$} &   \multicolumn{1}{c}{$46.86^{+30.79}_{-24.86}$} &  \multicolumn{1}{c}{$44.81^{+30.79}_{-23.49}$} \\ \vr

20 & \multicolumn{1}{c}{$1.63^{+ 0.42}_{-0.36}$} &  \multicolumn{1}{c}{$1.56^{+0.41}_{-0.33}$}  &  \multicolumn{1}{c}{$11.16^{+ 1.97}_{-1.86}$}  & \multicolumn{1}{c}{$10.11^{+2.52}_{-2.29}$}  & \multicolumn{1}{c}{$12.09^{+ 1.27}_{-1.84}$}  & \multicolumn{1}{c}{$9.70^{+1.96}_{-1.86}$}  &  \multicolumn{1}{c}{$0.27^{+0.22}_{-0.17}$} & \multicolumn{1}{c}{$0.83^{+0.08}_{-0.13}$} &   \multicolumn{1}{c}{$49.82^{+30.33}_{-26.68}$} &  \multicolumn{1}{c}{$44.12^{+29.88}_{-22.81}$} \\ \vr

21 & \multicolumn{1}{c}{$1.42^{+ 0.37}_{-0.26}$} &  \multicolumn{1}{c}{$1.78^{+0.40}_{-0.41}$}  &  \multicolumn{1}{c}{$10.64^{+ 1.71}_{-1.54}$}  & \multicolumn{1}{c}{$12.26^{+1.86}_{-2.44}$}  & \multicolumn{1}{c}{$12.26^{+ 1.17}_{-1.71}$}  & \multicolumn{1}{c}{$11.83^{+1.37}_{-1.98}$}  &  \multicolumn{1}{c}{$0.26^{+0.21}_{-0.16}$} & \multicolumn{1}{c}{$0.82^{+0.08}_{-0.13}$} &   \multicolumn{1}{c}{$41.84^{+31.47}_{-22.12}$} &  \multicolumn{1}{c}{$38.42^{+30.56}_{-19.84}$} \\ \vr

22 & \multicolumn{1}{c}{$1.33^{+ 0.28}_{-0.21}$} &  \multicolumn{1}{c}{$1.47^{+0.42}_{-0.30}$}  &  \multicolumn{1}{c}{$10.11^{+ 1.51}_{-1.48}$}  & \multicolumn{1}{c}{$9.74^{+ 2.84}_{-2.29}$}  & \multicolumn{1}{c}{$11.83^{+ 1.40}_{-1.79}$}  & \multicolumn{1}{c}{$7.91^{+1.87}_{-1.60}$}  &  \multicolumn{1}{c}{$0.26^{+0.21}_{-0.17}$} & \multicolumn{1}{c}{$0.83^{+0.09}_{-0.16}$} &   \multicolumn{1}{c}{$43.21^{+31.93}_{-22.81}$} &  \multicolumn{1}{c}{$47.32^{+30.56}_{-25.09}$} \\ \vr

23 & \multicolumn{1}{c}{$2.01^{+ 0.42}_{-0.42}$} &  \multicolumn{1}{c}{-}  &  \multicolumn{1}{c}{$11.85^{+ 1.68}_{-1.74}$}  & \multicolumn{1}{c}{-}  & \multicolumn{1}{c}{$12.36^{+ 1.10}_{-1.67}$}  & \multicolumn{1}{c}{$-$}  &  \multicolumn{1}{c}{$0.33^{+0.24}_{-0.21}$} & \multicolumn{1}{c}{$-$} &   \multicolumn{1}{c}{$46.86^{+30.33}_{-24.86}$} &  \multicolumn{1}{c}{$-$} \\ \vr

24 & \multicolumn{1}{c}{$1.40^{+ 0.33}_{-0.24}$} &  \multicolumn{1}{c}{$1.76^{+0.40}_{-0.40}$}  &  \multicolumn{1}{c}{$10.52^{+ 1.65}_{-1.51}$}  & \multicolumn{1}{c}{$12.14^{+1.97}_{-2.64}$}  & \multicolumn{1}{c}{$12.32^{+ 1.14}_{-1.67}$}  & \multicolumn{1}{c}{$11.65^{+1.49}_{-2.14}$}  &  \multicolumn{1}{c}{$0.27^{+0.22}_{-0.17}$} & \multicolumn{1}{c}{$0.80^{+0.09}_{-0.15}$} &   \multicolumn{1}{c}{$45.95^{+31.93}_{-24.63}$} &  \multicolumn{1}{c}{$50.74^{+30.56}_{-27.14}$} \\ \vr

25 & \multicolumn{1}{c}{$1.42^{+ 0.35}_{-0.26}$} &  \multicolumn{1}{c}{$1.40^{+ 0.33}_{-0.25}$}  &  \multicolumn{1}{c}{$9.79^{+ 1.74}_{-1.59}$}  & \multicolumn{1}{c}{$9.27^{+1.68}_{-2.00}$}  & \multicolumn{1}{c}{$12.34^{+ 1.13}_{-1.73}$}  & \multicolumn{1}{c}{$11.90^{+1.42}_{-2.05}$}  &  \multicolumn{1}{c}{$0.27^{+0.23}_{-0.17}$} & \multicolumn{1}{c}{$0.72^{+0.12}_{-0.19}$} &   \multicolumn{1}{c}{$50.51^{+30.33}_{-27.14}$} &  \multicolumn{1}{c}{$49.82^{+30.56}_{-26.68}$} \\ \vr

26 & \multicolumn{1}{c}{$1.48^{+ 0.50}_{-0.31}$} &  \multicolumn{1}{c}{$1.88^{+0.43}_{-0.42}$}  &  \multicolumn{1}{c}{$10.17^{+ 1.83}_{-1.83}$}  & \multicolumn{1}{c}{$13.59^{+1.51}_{-2.29}$}  & \multicolumn{1}{c}{$11.88^{+ 1.42}_{-2.13}$}  & \multicolumn{1}{c}{$12.04^{+ 1.18}_{-1.65}$}  &  \multicolumn{1}{c}{$0.35^{+0.27}_{-0.22}$} & \multicolumn{1}{c}{$0.76^{+0.09}_{-0.13}$} &   \multicolumn{1}{c}{$37.51^{+30.56}_{-19.16}$} &  \multicolumn{1}{c}{$37.05^{+30.56}_{-18.93}$} \\ \vr

27 & \multicolumn{1}{c}{$1.38^{+ 0.30}_{-0.24}$} &  \multicolumn{1}{c}{$1.42^{+ 0.37}_{-0.25}$}  &  \multicolumn{1}{c}{$8.46^{+ 1.51}_{-1.30}$}  & \multicolumn{1}{c}{$7.18^{+2.20}_{-1.39}$}  & \multicolumn{1}{c}{$12.02^{+ 1.30}_{-1.84}$}  & \multicolumn{1}{c}{$8.27^{+2.09}_{-1.45}$}  &  \multicolumn{1}{c}{$0.34^{+0.25}_{-0.22}$} & \multicolumn{1}{c}{$0.76^{+0.13}_{-0.25}$} &   \multicolumn{1}{c}{$42.07^{+31.47}_{-22.12}$} &  \multicolumn{1}{c}{$35.68^{+28.96}_{-17.79}$} \\

\hline

Combined & \multirow{2}{*}{$1.36^{+ 0.07}_{-0.07}$}& \multirow{2}{*}{$1.21^{+ 0.05}_{-0.05}$} &\multirow{2}{*}{$9.9^{+ 0.4}_{-0.4}$}&\multirow{2}{*}{$7.0^{+ 0.4}_{-0.4}$} & \multirow{2}{*}{$12.2^{+ 0.4}_{-0.5}$}& \multirow{2}{*}{$7.2^{+0.3}_{-0.4}$} &  \multirow{2}{*}{$0.16^{+0.05}_{-0.04}$} & \multirow{2}{*}{$0.84^{+0.01}_{-0.01}$} &  \multirow{2}{*}{$26.9^{+ 5.3}_{-4.4}$}  &  \multirow{2}{*}{$27.8^{+ 6.7}_{-4.3}$} \\

burst    &  &                   &                   &                   &  \\ 

\hline

\end{tabular}%

\caption{Medians and 68\% credible intervals of the mass, radius, distance, observer’s inclination, and column density for each burst and when information from all of the bursts is combined, for the \texttt{Bkg free} and \texttt{Bkg constrained (con)} cases. 
}
\label{tab:each_post}
\end{minipage}

\end{table*}

\subsection{Burst and burst oscillation properties }

\begin{figure*}%[h]
    \centering
    
    \includegraphics[width=1.8\columnwidth]{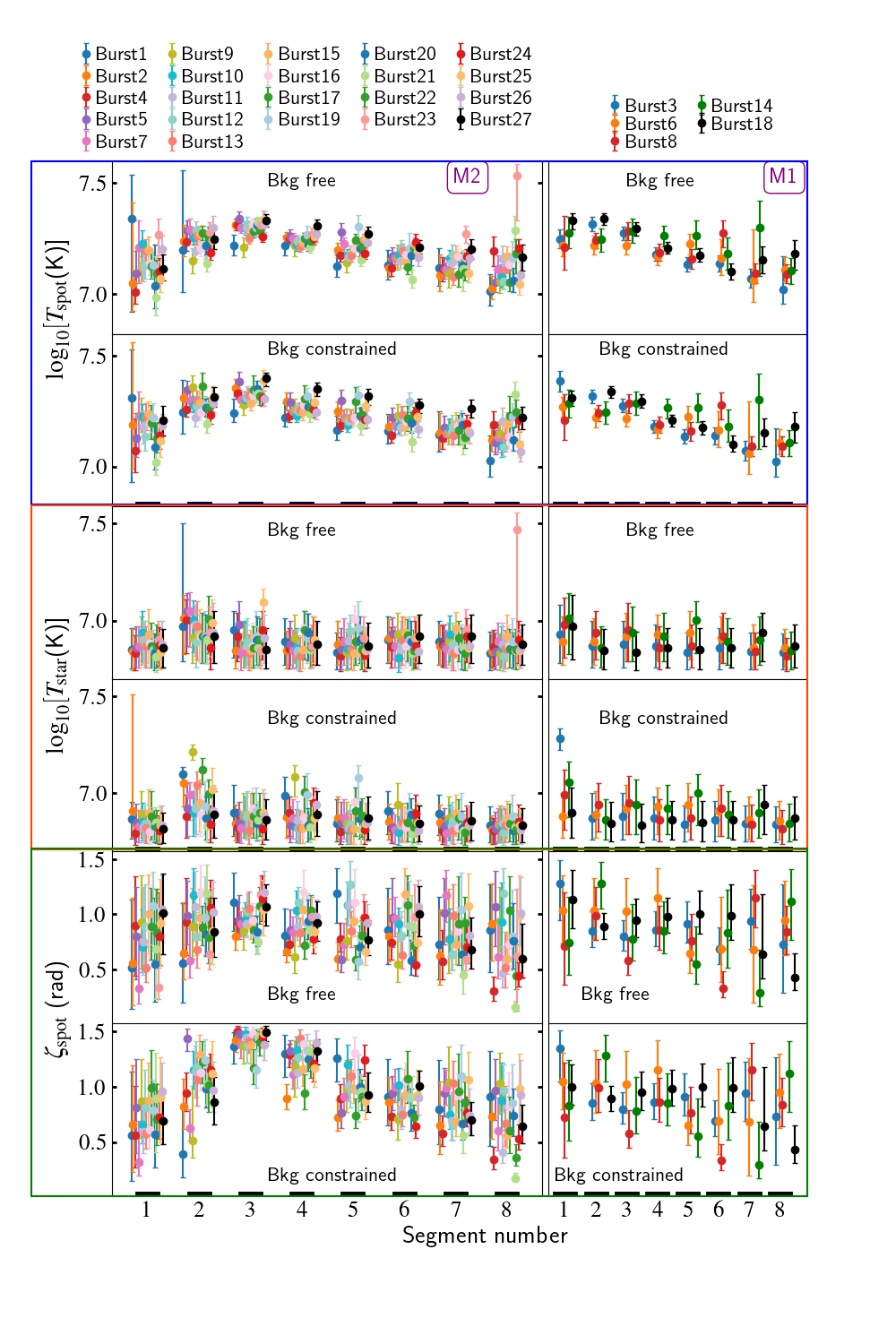}
      \caption{Time evolution of the hot spot temperature, star temperature, and angular radius of the hot spot. The left panel shows these parameters' evolution for bursts categorized as M2, while the right panel depicts bursts categorized as M1.}
    \label{fig:evolutions}
\end{figure*}

Given that the origins of the burst oscillations and the burning characteristics are not known a priori, we assume the hot spot temperature, angular radius, and stellar temperature are free parameters that can vary across segments, as mentioned in Section \ref{sec:models}. By adopting this approach, we aim to capture the variations in these parameters, which could offer insights into the poorly understood physics underlying the burst and oscillation properties. In Figure \ref{fig:evolutions}, we present the time evolution of the hot spot temperature, angular radius and stellar temperature or both the \texttt{Bkg free} and the \texttt{Bkg constrained} case. We plot the inferred medians, along with the 68\% credible intervals for each morphology (M1 and M2).

The models employed for inference allow for the possibility of the hot spot size and temperature as well as the rest of the star temperature to change throughout the burst, while the position of the hot spot remains fixed. This flexibility leads to a multitude of potential permutations. However, the results of the inference runs indicate a clear preference for a particular solution: one where the stellar temperature remains constant during the bursts, while the temperature and size of the hot spot vary. The stellar temperature is mostly consistent with being constant at the $1\sigma$ level whereas the hot spot temperature and size are not, especially in the \texttt{Bkg constrained} case.

\begin{figure}%[h]
    \centering
    
    \includegraphics[width=1.\columnwidth]{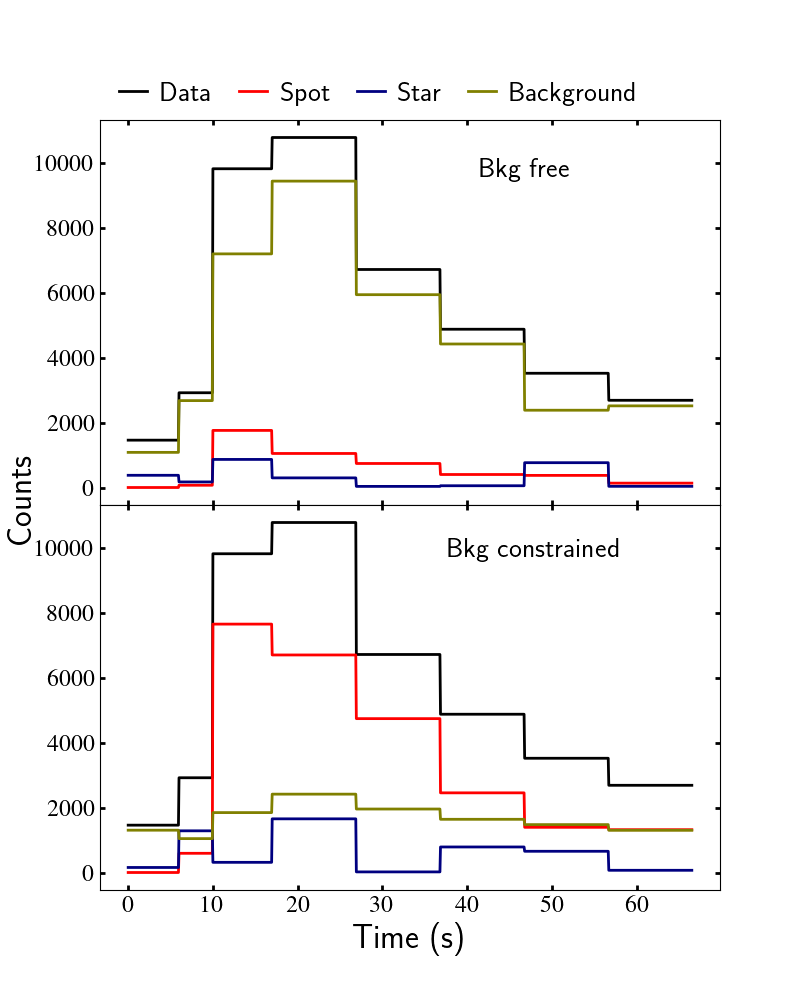}
      \caption{Temporal evolution in counts for Burst 1, showing contributions from the hot spot, the rest of the star, and the background. The counts are obtained using the maximum likelihood solution. In the \texttt{Bkg free} case, most of the burst photons originate from the background rather than any emitting component (hot spot \& star) on the stellar surface. However, once background constraints are applied (\texttt{Bkg constrained}), this situation changes and aligns more closely with the physical understanding of how bursts occur.}
    \label{fig:driver}
\end{figure}

Despite fluctuations in the hot spot temperature observed during the burst in the \texttt{Bkg free} scenario, it is evident that in that case the overall burst light curve is mostly determined by the changes in the background, as illustrated in Figure \ref{fig:driver}. Figure \ref{fig:driver} shows an example (for Burst 1, see Zenodo at \url{https://doi.org/10.5281/zenodo.8365643} for the remaining bursts) of the contribution from the hot spot, the star, and the background to the overall burst as a function of time. Both the hot spot and stellar contributions to the overall counts are negligible. While a dominating background solution is favored in \texttt{Bkg free}, it likely does not accurately reflect the true nature of a burst, as we discuss in Section \ref{sec:discussion}. Conversely, in the \texttt{Bkg constrained} case, the primary factors influencing the overall burst light curve are the variations in hot spot temperature and size, with the exception of five bursts (Bursts 1,6,8,9,14) where the background contribution remains high (see Zenodo at \url{https://doi.org/10.5281/zenodo.8365643}). For these bursts this is due to a high background limit derived from elevated pre-burst levels (see Section \ref{sec:bkg}).

\begin{figure}%[h]
    \centering
    
    \includegraphics[width=1.\columnwidth]{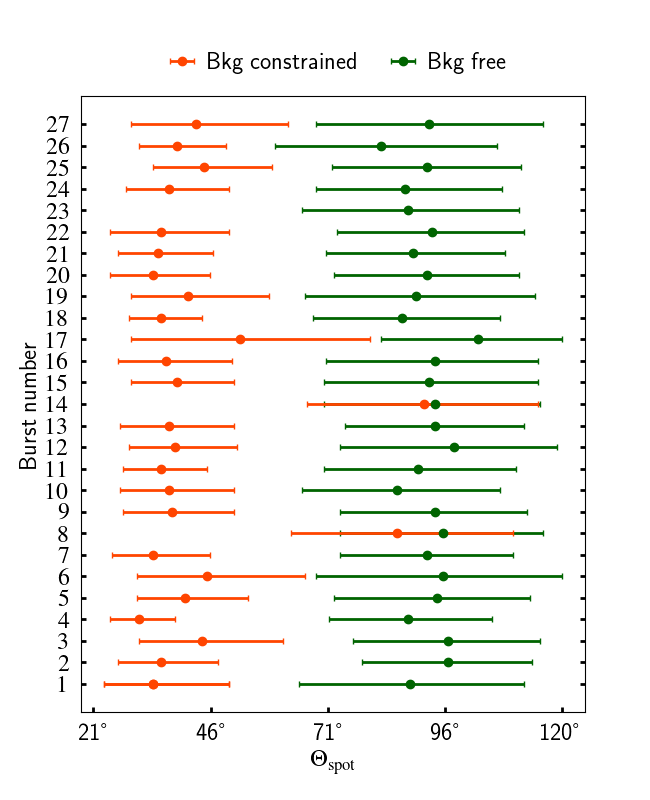}
      \caption{Median and the 68\% credible interval of the co-latitude of the centre of the hot spot for each burst. In the \texttt{Bkg constrained} case, the results are not shown for Burst 23 given that the run had to be halted due to high computational cost (see Section \ref{sec:result}).}
    \label{fig:theta}
\end{figure}

Both the \texttt{Bkg free} and \texttt{Bkg constrained} scenarios exhibit a preference for a large hot spot size. In the \texttt{Bkg free} case, the hot spots tend to be located away from the poles, typically in the vicinity of the equator. Conversely, in the \texttt{Bkg constrained} scenario, the hot spots are mostly located roughly halfway between the equator and the stellar northern rotational pole, as illustrated in Figure \ref{fig:theta}. Figure \ref{fig:theta} shows the median and the 68\% credible interval of the co-latitude of the centre of the hot spot for each burst for both the \texttt{Bkg free} and the \texttt{Bkg constrained} cases. The co-latitude of the centre of the hot spot for burst 8 and burst 14 is similar in both cases because they have particularly high inferred backgrounds. 

%% file: Sec3-Discussion.tex
\section{Discussion}\label{sec:discussion}

\subsection{Common parameters}

\subsubsection{Initial runs: \texttt{Bkg free} case}\label{sec:discussin_initial_runs}

\begin{figure}%[h]
    \centering
    
    \includegraphics[width=1.\columnwidth]{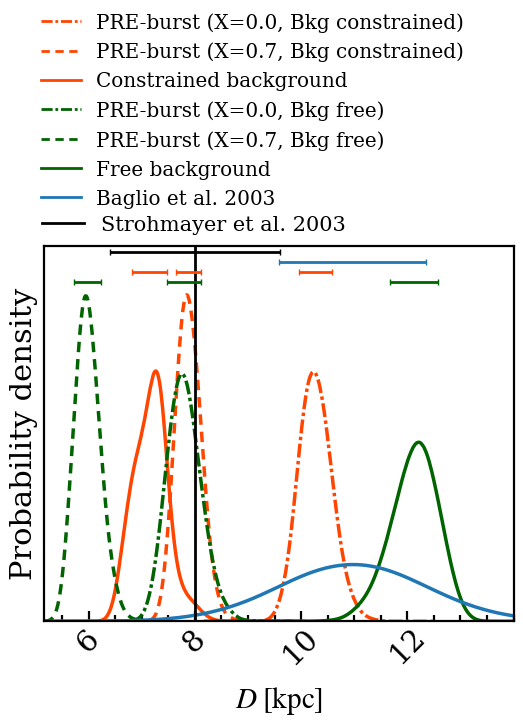}
      \caption{Posterior distributions of the distance. The color-coded horizontal segments are the 68\% credible intervals of each distribution symmetric in marginal posterior mass about the median. In the \citet{Strohmayer:2003} case, the horizontal segment is the 20\% uncertainty on the inferred distance. PRE-burst (X, "B") denotes that the distance distribution is computed using \texttt{Concord}, assuming a hydrogen fraction of X, along with the median of the mass and radius, and observer inclination distribution obtained under the "B" assumption, where "B" is either \texttt{Bkg free} or \texttt{Bkg constrained}.}
    \label{fig:distances}
\end{figure}

The inferred background in the \texttt{Bkg free} case is very high, dominating the overall burst count rate, while the contribution from the spot and the stellar component is almost negligible (see Figure \ref{fig:driver}). Such a scenario is inconsistent with current Type I X-ray burst models, where bursts arise from unstable thermonuclear burning of accreted fuel on the surface of neutron stars \citep[e.g.][]{Hansen:1975}, rather than simply from an increase in the persistent emission from the accretion disk/column.

The inferred distance for the \texttt{Bkg free} case is very large, $12.2^{+0.4}_{-0.5}$ kpc. It is, however, consistent with the distance inferred by \citet{Baglio:2012} (see Figure \ref{fig:distances}). Leveraging data obtained from the European Southern Observatory Very Large Telescope, \citet{Baglio:2012} conducted an analysis of the quiescent optical counterpart of \J,\ using multiband orbital phase-resolved photometry. Employing the Markov Chain Monte Carlo method to analyze and model the light curves, they constrained the distance to \J.\ Assuming a neutron star mass of 1.4 \msol, which is close to the inferred mass of \J\ in the \texttt{Bkg free} scenario, they arrived at a distance modulus of 15.2, equivalent to an estimated distance of approximately 11 kpc. However, the inferred distance in the \texttt{Bkg free} scenario does not align with the upper limit of 9.6 kpc deduced by \citet{Strohmayer:2003}. 

\citet{Strohmayer:2003} assumed a canonical star (similar to our mass and radius findings), with a hydrogen-rich photosphere, and used the peak luminosity of Burst 28 as the Eddington luminosity. When computing this upper limit, certain considerations, such as burst anisotropy, were not taken into account in their approach. To investigate how factors such as anisotropy and fuel composition impact the distance estimate, we employed a recently developed code \texttt{Concord} \citep{Galloway:2022}. \texttt{Concord}\footnote{\url{https://github.com/outs1der/concord}} offers a comprehensive approach to estimating Type I X-ray burster system parameters such as the distance while fully accounting for astrophysical uncertainties. In Figure \ref{fig:distances}, we present the expected distance distributions across various atmospheric compositions. These distributions are derived from the medians of the masses and radii and observer inclination distributions obtained under both the \texttt{Bkg free} and \texttt{Bkg constrained} cases.
The distance inferred in the \texttt{Bkg free} case is inconsistent with the distance distributions when considering Burst 28 as a PRE burst with the mass, radius, and inclination that we infer for this case.

Although the absence of an X-ray eclipse during the \J\ outburst implies that $\cos(i) >$ 0.2 \citep{Krauss:2005sj}, we refrained from enforcing this constraint in our prior on the observer inclination. Instead, we opted for a significantly broader prior distribution. In the \texttt{Bkg free} scenario, only half of the posterior samples are consistent with the absence of an eclipse.

Given these issues, it is clear that when the background is unconstrained, the preferred solution likely does not reflect correctly the properties of \J.\ In what follows we therefore discuss only the \texttt{Bkg constrained} case.

\subsubsection{\texttt{Bkg constrained} case}

When we constrain the background, we find a mass of $1.21^{+0.05}_{-0.05}$ \msol, a radius of $7.0^{+0.4}_{-0.4}$ km, with the NS located at a distance of $7.2^{+0.3}_{-0.4}$ kpc. The inferred radius is small compared to values inferred from other astrophysical and laboratory measurements \citep[see e.g.][]{Raaijmakers:2021,Huth22}.  While the median radius, marginalized over other parameters, for individual bursts, is typically greater than 8 km, a contrasting trend emerges when combining the information from all bursts. This discrepancy arises from the fact that the joint posterior distribution of mass, radius, and distance have a wide scatter for each burst, with regions of high probabilities around 10 km. Yet, the slopes of correlations among mass, radius, and distance are different, and the posteriors of different bursts intercept in regions of lower radii. Consequently, when combining results from all bursts, only regions within the radius space featuring smaller values exhibit high posteriors, thereby yielding the small inferred radius. The small inferred radius when all bursts are combined, coupled with the fact that the rms FA of the bolometric pulse of each burst especially for the first harmonic reaches lower values than that of the data, suggests that a single hot spot model might not be adequate to explain the burst oscillation properties.

In the \texttt{Bkg constrained} scenario, the background contribution to the burst is relatively low compared to the contribution from the hot spot. The bursts are predominantly dominated by photons emitted from the hot spot, while the stellar counts remain relatively low. This depiction aligns more closely with our current understanding of Type I X-ray burst physics. However, there are still five bursts where the number of counts from the background is either higher or nearly equal to the number of counts from the hot spot. This is due to how we define the start of the burst which led to a high pre-burst count rate for these bursts. Our simplistic assumption, that the upper limit on the background is the upper value of \texttt{fa }scaling factor times the pre-burst spectrum led to a big upper limit on the background, especially for faint bursts. Therefore, obtaining a better understanding of the background behavior during a burst is essential for future analyses.

Constraining the background leads to a consistent distance estimate obtained using \texttt{Concord} (see Figure \ref{fig:distances}) assuming a hydrogen-rich burst. However, this distance estimate cannot be compared directly to those reported by \citet{Baglio:2012} and \citet{Strohmayer:2003} since a 1.4 \msol was assumed in these works and we are now inferring a lower value. Employing the methodology of \citet{Strohmayer:2003}, we derive a distance of $7.7^{+1.5}_{-1.5}$ assuming a $1.21$ \msol NS, which aligns well with our inferred distance.

All samples for the observer's inclination agree with the absence of an eclipse \citep{Krauss:2005sj}. Furthermore, our findings are consistent (across the entire posterior distribution) with those of \citet{Krauss:2005sj}, where it was inferred from the magnitude limit of the optical counterpart that $\cos(i) \lesssim 0.93$.

Regarding the line-of-sight hydrogen column density, our results are higher than the commonly reported range of $(16-17)\times 10^{20}\mathrm{cm^{-2}}$, a prevalent finding within the existing literature concerning \J\ \citep[see e.g.,][]{Krauss:2005sj, Galloway:2020}. These line-of-sight hydrogen column densities are computed using \citet{Dickey:1990} HI column density\footnote{https://cxc.harvard.edu/toolkit/colden.jsp}. The line-of-sight hydrogen column density is also higher than $15^{+2.1}_{-1.9}\times 10^{20}\mathrm{cm^{-2}}$ obtained using the 3D interstellar absorption map \citep{Doroshenko:2024}\footnote{http://astro.uni-tuebingen.de/nh3d/nhtool}.

\subsection{Implications for Equation of State}

In the \texttt{Bkg constrained} case, where the physics aligns with our current understanding of Type I X-ray bursts mechanism, we found a $1.21^{+0.05}_{-0.05}$ \msol star with a radius of $7.0^{+0.4}_{-0.4}$ km, indicating a preference for softer EoSs.
The NICER collaboration has inferred the mass and radius of two NS. One has a high mass (about 2.0\msol), with tight mass constraints inferred from radio observations \citep{Cromartie:2020}, PSR J0740+6620 \citep{Miller:2021,Riley:2021, Salmi:2022}. The second, PSR J0030$+$0451 had no mass prior constraint. Initial analyses suggested a mass of about 1.4 \msol and a radius of about 13 km \citep{Miller:2019, Riley:2019}. Updated analysis, using a new NICER response and improved background constraints from XMM-Newton \citep{Vinciguerra:2024}, found a mode compatible with a canonical neutron star with a lower radius of $11.71^{+0.88}_{-0.83}$ km (68\% CI).

By fitting the X-ray burst cooling tail spectra of 4U 1702-429 with different models, \citealt{Nattila:2017} derived for one of the models considered -model $\mathcal{A}$-that 4U 1702-429 could be a canonical star with a radius of $\mathrm{R_{1.4}}=12.4^{+0.4}_{-0.4}$ km (68\% CI). Although model $\mathcal{A}$ was not the preferred model, it would be broadly consistent with the NICER results for the radius of a 1.4 \msol star.

The radius of a 1.4\msol star has also been derived through EoS inference. By integrating multiple pieces of information, including tidal deformability data from the gravitational wave event GW170817 \citep{GW170817} and measurements from both PSR J0030+0451 and PSR J0740+6620, and applying EoS inference techniques, both \citealt{Miller:2021} and \citealt{Raaijmakers:2021} arrived at similar results, indicating a radius of approximately $\mathrm{R_{1.4}} \approx 12$km. 
Radii measurements from tidal deformability from the gravitational wave event GW170817 also favor softer EoS \citep{GW170817_radius}.  Similar results have been obtained by other groups, see for example \citet{Huth22}, whose EoS analysis yields $\mathrm{R_{1.4}} = 12.01^{+0.78}_{-0.77}$ km.

All the above-mentioned results are broadly consistent with each other. However, the radius inferred for \J \ is notably smaller than the radius expected for a 1.2 \msol NS \citep[see e.g.][]{Raaijmakers:2021, Huth22}, which may imply limitations in our single hot spot model's ability to depict the data accurately. However, the possibility of such a small radius cannot be ruled out, especially when considering the presence of strong phase transitions in quantum chromodynamics. Such a phase transition would allow hybrid stars, in the second branch, with considerably smaller radii than the current constraints \citep{Christian:2022,Jie:2024}. While such a small radius remains plausible, no configuration has been found where a star this compact could form while simultaneously satisfying all other astrophysical constraints \citep{Christian:2024} as such a small star would require an exceptionally low transition pressure $(p_\mathrm{trans}<50 \mathrm{MeV/fm^3})$\footnote{Private discussion with Jan-Erik Christian.}.

\subsection{Burst and burst oscillation properties: \texttt{Bkg constrained} case}

The evolution of the star's temperature, the hot spot temperature, and the hot spot angular radius for each burst suggest that the burning that occurred on \J\ during its 2003 outburst was confined. This is mainly because photons originating from the hot spot dominate the burst emission, coupled with the fact that the hot spot temperature and the hot spot angular radius change over time while the stellar temperature remains constant. Remarkably the stellar temperature remains relatively low, about $(0.64\pm 0.16)$ keV throughout all bursts. The temperature is consistent with the X-ray spectral properties inferred by \citet{Krauss:2005sj} for \J\ during its outburst.

If the spot locations reflect the starting point of the bursts, then ignition mostly occurs halfway between \J's\ equator and its northern rotational pole. The flames then propagate and engulf about half of the star before dying out. Given the weak magnetic field expected for \J\, fast rotation is likely what halts the flames from spreading across the entire surface \citep{Cavecchi:2016}. However, the ignition location is challenging to reconcile with the hypothesis that ignition is more likely to take place at the equator, mainly due to the influence of rapid rotation, which reduces effective gravity in that area \citep{Spitkovsky:2002,Cooper:2007}. The possibility of channeled accretion leading to off-equatorial ignition in pulsars has been explored, but as yet no mechanism has been identified \citep{Goodwin:2021}.

 It is important to note that there are significant uncertainties associated with the inferred time-varying parameters. For example, the 99\% credible regions for the majority of the angular radii of each segment are nearly identical to the prior space. In \citet{kini:2023b}, the uncertainties associated with time-varying parameters were considerably lower. This was primarily due to the hierarchical evolution imposed on the time-varying parameters coupled with tight constraints on both the distance and the background. In addition to the lack of constraints on time-varying parameters in this work, the significant level of uncertainty can be attributed to several factors. Firstly, the quality of the available data might not be sufficient to fully capture the complex behavior of the hot spot during the burst. Secondly, the duration of each time segment could be too lengthy to detect subtle variations. As a result, drawing definitive conclusions about the evolution of temporally varying parameters within each burst becomes challenging based solely on this analysis. Given these limitations, it becomes apparent that obtaining improved data quality is desirable. Moreover, the necessity for more computing resources becomes evident for understanding the parameter evolutions during the burst.

\subsection{Comparison with previous analyses of \J}\label{sec:compare}

\citet{Bhattacharyya:2004pp} have modeled the TBOs of 22 bursts of \J\ using a single hot spot model as in this paper. However, there are some major differences in the analysis approach e.g.: \citealt{Bhattacharyya:2004pp} did not incorporate time variability in their analysis, nor did they consider the effects of the interstellar medium, and the temperature of the hot spot was fixed at 2 keV. Moreover, bursts were stacked during the modelling procedure, they considered a spherical star with frame dragging, and the entire analysis was conducted assuming two EoS. Their analysis yielded constraints on the compactness, suggesting  $M/R_\mathrm{eq} < 1/4.2 \approx 0.238$ at a 90\% confidence level\footnote{As in Figure \ref{fig:combed_MR}, we set G=c=1.}. They also determined the center of the hot spot ($\Theta_\mathrm{spot}$) to be within the range of $60^{\circ}$ to $139^{\circ}$ (with regions of $\Theta_\mathrm{spot}<50^{\circ}$ being deemed highly unlikely), at a 90\% confidence level. Moreover, they identified a high probability for $i> 22^{\circ}$. Our results also align with their lower limit on the observer inclination but only approximately 5\% of the inferred posterior mass of compactness is below their compactness upper limit for the \texttt{Bkg constrained} case. Also, we observed a discrepancy regarding the center of the hot spot; our findings suggest that for most bursts, the most probable location lies between $20^{\circ}$ and $50^{\circ}$, whereas \citealt{Bhattacharyya:2004pp} identified this range as highly unlikely.
Another study that has constrained the properties of \J\ is that conducted by \citet{Wang:2017}. Through phase-resolved spectroscopy of its 2003 outburst, they derived a mass of $2.0^{+0.7}_{-0.5}$\msol with their lower limit being higher than our 68\%  CI upper limit. Assuming a pulsar mass of 1.4\msol, they concluded that $i \gtrsim 48^{\circ}$. Such constraint on the observer inclination does not align with the inferred posterior for the inclination in the \texttt{Bkg constrained} case, where all our samples have $i<48^{\circ}$. It is crucial to highlight that their assumption of the mass does not align with the mass inferred under \texttt{Bkg constrained} case. Also,  the prior space for $i$\footnote{For the prior on the observer inclination, they have $\cos(i)\sim\mathcal{U}(\cos(35^{\circ}),\cos(78^{\circ}))$.} assumed in their analysis inherently excludes our most probable solutions.

Using the accretion-powered pulsations (not the TBOs), \citet{Leahy:2009} have also inferred properties of \J. By fitting the bolometric pulse profiles of two bands (2-3 keV, and 7-9 keV) and using a single hot spot model and 23 days of observation from June 5–27, 2003, and excluding the X-bursts, the analysis mostly yielded high inferred mass and stiff equations of state solutions. Our analysis however favours a soft EoS solution.

\subsection{Improvements to this analysis}\label{sec:improvement}

There are a couple of aspects that our analysis has not thoroughly explored, some due to limited computing resources, and some for the lack of knowledge of the underlying physics. It would be beneficial to delve into these aspects in future research.

First, we assume that the atmosphere of \J\ is that of a thermonuclear burster. However it is possible that such an atmosphere model might not correctly describe the physical conditions of \J,\ if the pulsations originate from somewhere in the atmosphere. Second, we presumed that the atmosphere composition of \J\ has solar abundance.  This is mainly because the burst morphology strongly suggests a burning of hydrogen and helium \citep{Galloway:2008}. However, we cannot rule out the possibility that the atmospheric composition could be different\footnote{For \J,\ pure helium burning is highly improbable, since if it were to occur, the resulting burst shapes are expected to differ significantly from those observed in the 2003 bursts.}. It is also possible that the fraction of heavier elements in the atmosphere composition is different from what we currently assume (\texttt{Z = 0.0134}). Investigating how variations in atmospheric composition might influence the inferred parameters, particularly the mass and radius is therefore desirable. Yet, we note that given the data quality of the bursts, the sensitivity of our results to changes in atmosphere composition through pulse profile modelling might be limited.

We also employed a single uniform temperature hot spot to model the observed oscillations in the burst light curves. No discernible anomalies within the residuals or in the residual distributions indicate inadequate model performance. Nevertheless, the challenge of fitting the harmonic contents of the data with a single hot spot, along with the small inferred radius, hints at potential systematic bias. We could explore the impact of an additional constraint, ensuring that both the rms FA of the bolometric pulse, for both the fundamental and the first harmonic, match that of the data during sampling. Also given that \J\ is an AMXP, it exhibits pulsations arising from the accretion hot spot(s) as well. It is highly plausible that the hot spot causing burst oscillations is distinct from those generating accretion pulsations \citep[see e.g.][]{Watts:2008,Cavecchi:2022}. Therefore, the observed pulsations could potentially be a combined result of these hot spots: a burst oscillation hot spot and one or two (depending on whether the star is assumed to be entirely visible or not) hot spots due to the accretion. While during the peak of the burst, the contribution from accretion pulsations is supposed to be negligible due to the dominance of burst oscillations \citep[][]{Watts:2005}, this might not hold true at the burst's start and end. Notably, the accretion hot spot's temperature is estimated to be $\sim 1$ keV \citep{Poutanen:2003, Salmi:2018, Das:2022}, which coincidentally aligns with the temperature we determined for the burst oscillation at the tail and beginning of the bursts. This is a possible explanation for the increase in the hot spot temperature observed in the tail of the bursts. Ideally, exploring scenarios involving multiple hot spots would be insightful and potentially yield much higher rms FA of the first harmonic. %However, this endeavour might introduce unnecessary complexities into the model given the current data quality.

Furthermore, given that \J\ is an AMXP, it is surrounded by both an accretion column and an accretion disk. It is possible that the influence of both these components could impact the observed spectrum but in this work, we overlooked both of them. The effects of the accretion disk should, in theory, be less pronounced, as they would predominantly affect the lower energy photons where RXTE is less sensitive. Nevertheless, it is desirable to scrutinize the effects of this component in more detail. Furthermore, we assumed that the entire star remains visible throughout the burst given that the burst might push away the accretion disk \citep[see e.g.][]{Fragile:2020}, rendering both hemispheres of the star visible. But such a scenario is plausible during PRE bursts, which the bursts analyzed in this analysis are not. Moreover, possibly obscuration from accretion was overlooked. 

Finally, we kept the phase and co-latitude of the hot spot fixed across all time segments. Nevertheless, previous analysis of \J\ has shown a phase drift of up to 10\% for some bursts \citep{Cavecchi:2022}. Although we checked that this phase drift does not significantly impede parameter inference for synthetic data with $10^5$ counts, ignoring this phase drift might pose issues when combining bursts. Ideally, we should leave both the phase of the hot spot and the co-latitude of the hot spot free for each segment. However, this would lead to extra computational time.

\section{Conclusion}\label{sec:conclusion}

We have conducted a comprehensive analysis of the AMXP and TBO source \J\ to infer its properties through the PPM of burst oscillations observed during its 2003 outburst. To achieve this, we employed a state-of-the-art PPM technique to derive key parameters such as the mass, radius, distance, and observer's inclination of \J.

Our analysis yielded the following results for \J:\ a mass of $1.21^{+0.05}_{-0.05}$ \msol, a radius of $7.0^{+0.4}_{-0.4}$ km, and a distance of $7.2^{+0.3}_{-0.4}$ kpc. Our result favors soft EoS for NSs. However, the relatively small radius inferred and the lack of harmonic content in the bolometric pulse may stem from potential systematic errors originating from our limited understanding of burst oscillation mechanisms. Therefore, further research to enhance our understanding of burst oscillation origins would be invaluable. Furthermore, while a previous analysis \citep{Bhattacharyya:2004pp} used a single uniform temperature hot spot model for burst oscillations, our analysis indicates potential shortcomings in this model once time variability is fully taken into account, highlighting the necessity for alternative models. 

Moreover, potential systematic biases originating from other various modelling assumptions made in this paper cannot be ruled out. Given that \J\ is an AMXP, modelling the accretion pulsations through PPM, investigating burst properties as for IGR J17498-2921 \citep{Galloway:2024} or SAX J1808.4-3658 \citep{Goodwin:2021, Galloway:2024} cor applying the direct cooling tail method \citep[see e.g.][]{Nattila:2017, Molkov:2024} would be valuable. By doing so, we can cross-check results and mitigate potential biases associated with PPM of TBO sources. Additionally, the correlation between the mass, radius, and distance varies for each burst, leading to the small inferred radius when combining the information from many bursts. Independent constraints on the distance would significantly improve future analyses of this source.

With about 2 million total counts, the uncertainties on both the mass and radius are about 10\% (using the 68\% CI). If \J-like bursts would be observed with proposed large-area X-ray spectral-timing telescopes like the enhanced X-ray Timing and Polarimetry mission (e-XTP) \citep[e.g.,][]{Zhang:2019, Watts:2019_extp}, or the Spectroscopic Time-Resolving Observatory for Broadband Energy X-rays (STROBE-X) \citep[e.g.,][]{Ray:2019}, this would result in much larger number of counts collected, hence reducing the uncertainties to only a few percent. This underscores the significance of modelling burst oscillation sources.

Our analysis also delved into the temporal evolution of hot spot temperatures, angular radii, and stellar temperatures during the bursts. While we observed marginal variation during the bursts of the hot spot temperature and the hot spot angular radius, the stellar temperature remained stable. This suggests a preference for confined burning where the flames initially spread across half the stellar surface and then stall. 

However, it is important to note that the quality of the available data, the number of each time segment used, and the absence of independent constraints on certain parameters, such as distance and background, have made it challenging to draw definitive conclusions regarding the true evolution of the time-dependent parameters. To gain deeper insights into the poorly understood physics underlying the burst and burst oscillations, improved data quality, increased computational resources, independent distance measurements and better knowledge of the atmosphere composition are highly desirable.

Additionally, although trying different background constraints yielded similar results in terms of mass, the inferred radius, distance, and observer inclination exhibited significant disparities. This underscores the importance of independent constraints on both distance and observer inclination to gain a better understanding of the poorly constrained background behavior during a Type I X-ray burst.

In sum, this study has provided insights into probable properties of \J\ and underscores the challenges and opportunities inherent in inferring TBOs sources properties through PPM. Future observations and modelling efforts will undoubtedly continue to enhance our understanding of NSs and the fundamental physics governing their behavior.

%% file: Sec4-Appendix.tex
%\appendix

\section{Appendix }\label{sec:appendix}

In Table \ref{tab:duration}, we summarise the duration of each time segment for a given burst.
\begin{table*}
\centering
    \resizebox{\textwidth}{!}{%
    \begin{tabular}{|c|c|c|c|c|c|c|c|c|}
    \hline
    Burst & \multicolumn{8}{c}{Duration (in s)} \\ \hline
    & Segment 1 & Segment 2 & Segment 3 & Segment 4 & Segment 5 & Segment 6 & Segment 7 & Segment 8 \\ \hline
    1 & 6.00 & 4.00 & 7.00 & 9.92 & 9.92 & 9.92 & 9.92 & 9.92 \\
    2 & 3.00 & 2.00 & 10.00 & 18.35 & 18.34 & 18.34 & 18.35 & 18.36 \\
    3 & 1.00 & 6.00 & 8.00 & 17.20 & 17.20 & 17.20 & 17.20 & 17.20 \\
    4 & 5.00 & 2.50 & 8.50 & 22.20 & 22.19 & 22.19 & 22.19 & 22.19 \\
    5 & 4.50 & 2.50 & 8.00 & 15.12 & 15.12 & 15.11 & 15.11 & 15.12 \\
    6 & 2.50 & 4.50 & 8.00 & 9.80 & 9.79 & 9.79 & 9.79 & 9.80 \\
    7 & 8.00 & 3.50 & 13.50 & 16.35 & 16.35 & 16.34 & 16.34 & 16.36 \\
    8 & 1.20 & 6.80 & 8.00 & 11.85 & 11.85 & 11.84 & 11.84 & 11.85 \\
    9 & 1.50 & 2.50 & 9.00 & 11.20 & 11.19 & 11.19 & 11.19 & 11.20 \\
    10 & 1.50 & 1.80 & 8.20 & 15.65 & 15.65 & 15.65 & 15.64 & 15.64 \\
    11 & 3.50 & 2.00 & 9.00 & 17.60 & 17.59 & 17.60 & 17.60 & 17.60 \\
    12 & 2.50 & 3.00 & 9.00 & 19.10 & 19.10 & 19.09 & 19.09 & 19.10 \\
    13 & 4.50 & 3.00 & 10.00 & 18.25 & 18.24 & 18.25 & 18.24 & 18.25 \\
    14 & 1.40 & 4.10 & 8.00 & 6.19 & 6.19 & 6.19 & 6.19 & 6.20 \\
    15 & 1.00 & 1.70 & 8.80 & 15.05 & 15.05 & 15.04 & 15.04 & 15.05 \\
    16 & 4.00 & 3.70 & 8.80 & 15.75 & 15.75 & 15.74 & 15.74 & 15.75 \\
    17 & 1.50 & 2.00 & 8.00 & 15.40 & 15.40 & 15.40 & 15.40 & 15.40 \\
    18 & 0.50 & 6.00 & 8.00 & 15.90 & 15.90 & 15.90 & 15.89 & 15.90 \\
    19 & 4.00 & 3.49 & 6.99 & 7.04 & 7.04 & 7.04 & 7.04 & 7.05 \\
    20 & 6.50 & 4.00 & 9.00 & 18.35 & 18.35 & 18.34 & 18.35 & 18.35 \\
    21 & 6.50 & 4.00 & 10.00 & 30.45 & 30.45 & 30.45 & 30.45 & 30.44 \\
    22 & 6.50 & 2.50 & 11.50 & 21.82 & 21.82 & 21.82 & 21.82 & 21.82 \\
    23 & 3.50 & 4.00 & 10.00 & 12.36 & 12.35 & 12.35 & 12.35 & 12.36 \\
    24 & 5.99 & 3.40 & 10.10 & 17.20 & 17.20 & 17.19 & 17.18 & 17.21 \\
    25 & 4.00 & 2.40 & 7.10 & 15.42 & 15.42 & 15.41 & 15.40 & 15.42 \\
    26 & 2.00 & 2.00 & 9.50 & 21.25 & 21.25 & 21.25 & 21.25 & 21.25 \\
    27 & 6.50 & 4.99 & 10.00 & 22.85 & 22.84 & 22.84 & 22.84 & 22.85 \\ \hline
        
    \end{tabular}
    }
    \caption{Duration of each burst segment.}
    \label{tab:duration}
\end{table*}

In Table \ref{tab:core_hours}, we show the run time for each burst for the \texttt{Bkg free} and \texttt{Bkg constrained} cases.

\begin{table}

\centering
\resizebox{1\columnwidth}{!}{%

\begin{tabular}{lcc}

\hline \hline

\multirow{2}{*}{Burst}  & \multicolumn{2}{c}{\multirow{2}{*}{CPU core-hour ($\times 10^{4}$)}} \\
                  & \multicolumn{2}{l}{}                  \\
                  \cline{2-3}
                  & \texttt{Bkg free}            & \texttt{Bkg constrained}    \\

\hline

 1 & 2.19   &  9.79 \\
  
 2 & 5.01  & 24.29  \\

 3 & 6.57  & 36.42 \\

 4 & 9.97 & 16.76 \\

 5 & 5.60  & 15.39 \\

 6 & 4.37  & 11.72  \\ 

 7 & 12.63  & 34.42\\

 8 & 7.01 &  10.59 \\

 9 & 5.25  & 23.71  \\

10 & 10.93  & 24.43\\

11 & 10.84  & 10.73  \\

12 & 10.40  & 11.36  \\

13 & 9.38  & 13.10 \\

14& 10.49 &  3.61 \\

15& 10.88  & 11.54 \\ 

16 & 7.08 & 20.63  \\ 

17 & 7.56  & 24.56  \\ 

18 & 10.05 & 11.26  \\

19 & 6.37 & 15.87 \\

20 & 8.67  & 12.11 \\
21 & 8.73  & 40.19  \\

22 & 12.75  & 42.69\\

23 & 25.35  & \textbf{40.25} \\

24 & 9.91 & 14.81\\

25 & 0.76 & 8.04 \\

26 & 9.68  & 6.51 \\

27 & 9.09 & 17.16 \\
\hline
Total & 237.52 & 511.94\\
\hline 
\end{tabular}
}
\caption{Core-hours spent on runs for each burst. Burst 23 in the \texttt{Bkg constrained} case had to be stopped due to resource constraints, having already consumed $40.25\times 10^{4}$ core-hours. }
\label{tab:core_hours}
\end{table}

\begin{figure}%[h]
    \centering
    \includegraphics[width=1.\columnwidth]{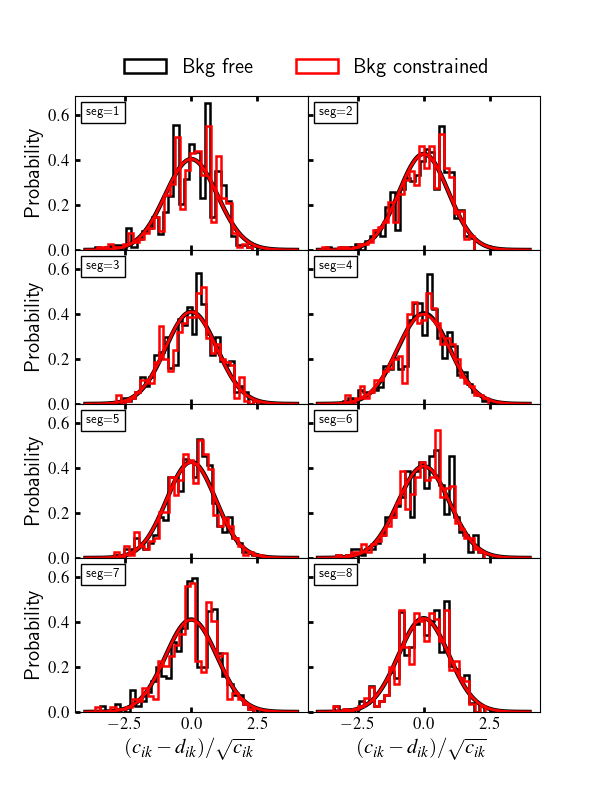}
      \caption{Histograms of the residuals for Burst 1 for \texttt{Bkg free} and \texttt{Bkg constrained} for all the segments. The residuals are the difference between the model counts (for the maximum likelihood solution) and the data counts, normalized by the model count counts in each instrument energy channel and phase bin. The solid lines are the Gaussian curves that fit best the histograms.}
    \label{fig:residuals_distibutions}
\end{figure}

In Figure \ref{fig:residuals_distibutions}, we show the distribution of the residuals for Burst 1 for \texttt{Bkg free} and \texttt{Bkg constrained} for all the segments. The residuals are the difference between the model counts (for the maximum likelihood solution) normalized by the model count counts in each instrument energy channel and phase bin. The distributions of the residuals for the remaining bursts are available on Zenodo (\url{{https://doi.org/10.5281/zenodo.8365643}}). There is no deviation from the overall expected Gaussian distribution which could hint at a shortcoming of the single hot spot model.

\begin{figure*}%[h]
    \centering
    \includegraphics[width=2.\columnwidth]{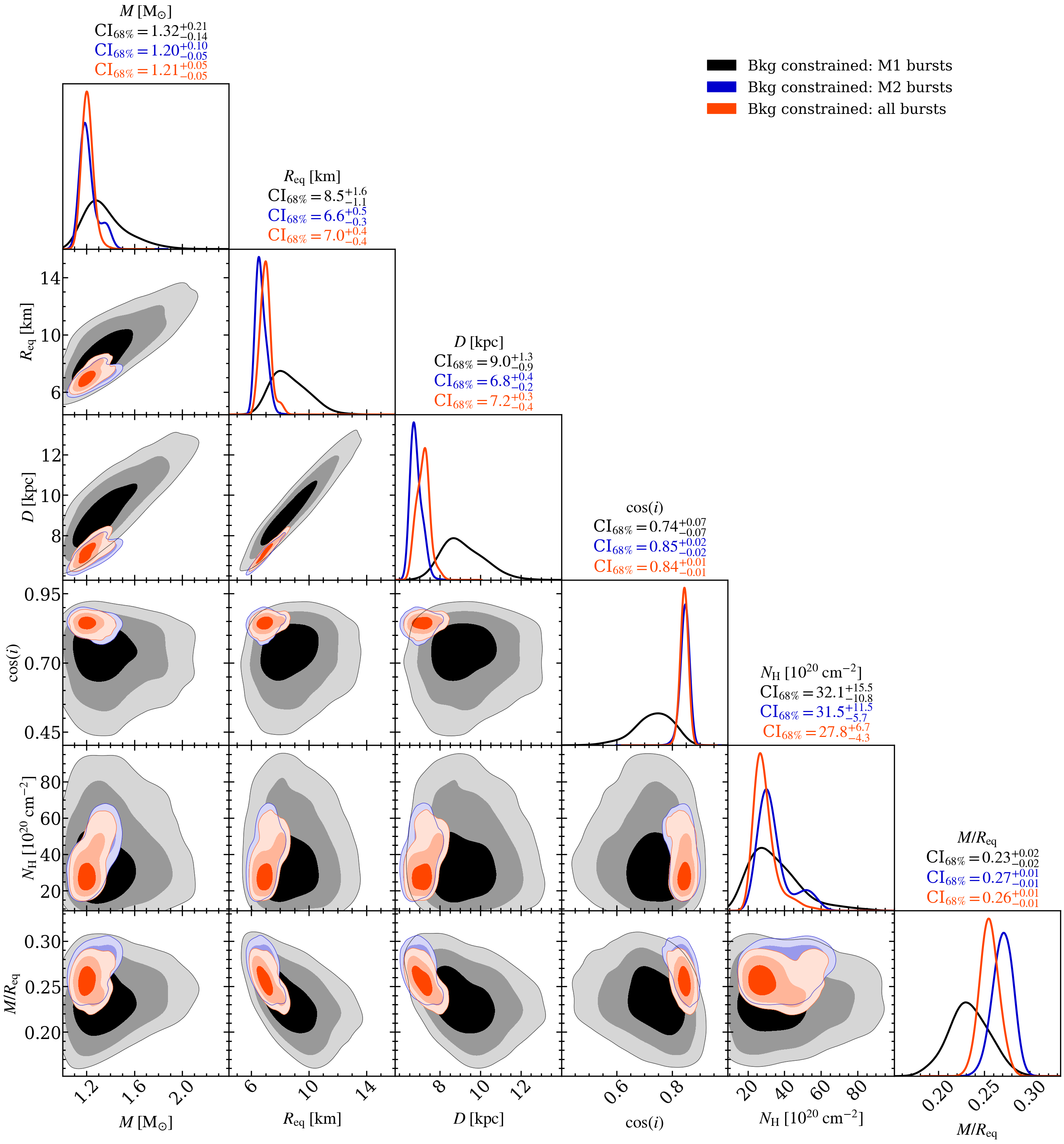}
      \caption{Combined posterior distributions of mass, radius, distance, observer inclination, column density, and compactness obtained using different \texttt{Multinest} live points. The two-dimensional posterior distributions are, from most opaque to least the 68\%, 95\%, and 99\% posterior credible region.}
    \label{fig:subset}
\end{figure*}

Figure \ref{fig:subset} shows the combined posteriors for each subset (M1 and M2), along with the entire set of bursts in the \texttt{Bkg constrained} case. Both the M1 and M2 bursts favor regions with low masses and radii. The combined posteriors using the entire set of bursts  are predominantly influenced by the M2 subset, as it contains a higher number of bursts (21 compared to 5 in the M1 subset).

\begin{figure*}%[h]
    \centering
    \includegraphics[width=2.\columnwidth]{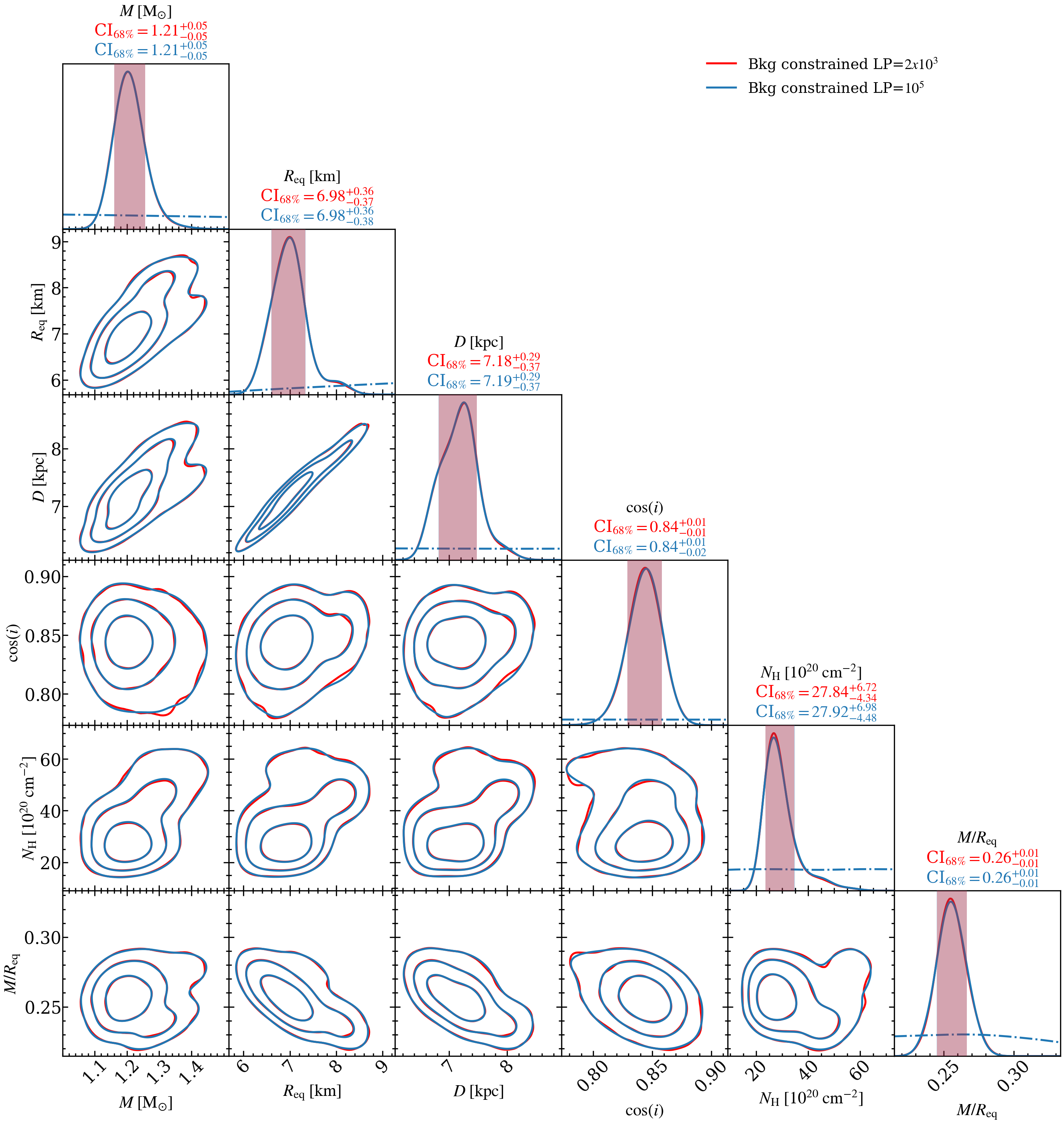}
      \caption{Combined posterior distributions of mass, radius, distance, observer inclination, column density, and compactness obtained using different \texttt{Multinest} live points. The two-dimensional contours are the 68\%, 95\%, and 99\% posterior credible region. The solid (dash-dotted) lines along the diagonal represent the marginalized posterior (prior) distribution of each parameter. The vertical bands are the inferred 68\% credible intervals.}
    \label{fig:compre_LP}
\end{figure*}

In Figure \ref{fig:compre_LP}, we show the combined posterior distributions of mass, radius, distance, observer inclination, column density, and compactness for \texttt{Bkg constrained} case. The posteriors were obtained using different \texttt{Multinest} live points. Increasing the number of live points by a factor of 50 yields the same results.

In Table \ref{tab:settings}, we provide a summary of the \texttt{Multinest} settings utilized in each inference run for individual bursts, as well as for the combined burst run. 
\begin{table}
 \centering
 \begin{minipage}{1.\columnwidth}
    \resizebox{\columnwidth}{!}{%
\begin{tabular}{lc}
\hline \hline
 Variable & Value \\
\hline
\texttt{multimodal}               & \texttt{False} \\
\texttt{n\_clustering\_params}    & \texttt{None}  \\
\texttt{n\_iter\_before\_update}  & \texttt{None } \\
\texttt{n\_live\_points}          & 2000  \\
\texttt{sampling\_efficiency}     & 0.1 \\
\texttt{const\_efficiency\_mode}  & \texttt{False} \\
\texttt{evidence\_tolerance}      & 0.1   \\
\texttt{max\_iter}                & -1    \\
\hline

\end{tabular}
}

\caption{\texttt{Multinest} settings used during the inference run of each burst as well as the combined burst runs, except for the combined burst we set \texttt{multimodal :True} }
    \label{tab:settings}
\end{minipage}
\end{table}

We have compiled in Table \ref{tab:live_points} the Maximum A Posteriori values corresponding to each mode discovered for the various choices of \texttt{Multinest} live points in the combined burst run.
% \begin{table*}
%  \centering
%  \begin{minipage}{2.\columnwidth}
%     \resizebox{\columnwidth}{!}{%
% \begin{tabular}{ccccccc}
% \hline \hline
%  Parameter & $M$ (\msol) & $R_\mathrm{eq}$ (km) & $D$ (kpc) & $\cos(i)$ & $N_H$ ($10^{20}\mathrm{cm}^{-2}$) & Log-Evidence \\
% \hline
%  \multicolumn{7}{c}{\textbf{$2\times10^3$ live points}} \\
% \hline
%  Mode1 & 1.23 & 7.3 & 7.3 & 0.85 & 27.79 & -2256.57$\pm$ 0.09 \\
%  Mode2 & 1.25 & 7.9 & 7.8 & 0.84 & 42.53 & -2308.18 $\pm$ 0.94 \\
%  Mode3 & 1.42 & 7.7 & 7.6 & 0.83 & 53.69 & -2314.70 $\pm$ 0.70 \\
% \hline
% \multicolumn{7}{c}{\textbf{$10^5$ live points}} \\
% \hline
% Mode1 & 1.23 & 7.4 & 7.6 & 0.85 & 26.14 & -2258.19 $\pm$ 0.12 \\
% Mode2 & 1.31 & 8.2 & 8.0 & 0.85 & 45.69 & -2309.25$\pm$ 0.13 \\
% Mode3 & 1.32 & 7.1 & 7.1 & 0.82 & 56.15 & -2314.86 $\pm$ 0.10 \\
% Mode4 & 1.29 & 8.0 & 7.9 & 0.86 & 46.74 & -2336.74$\pm$ 0.56 \\
% Mode5 & 1.35 & 7.2 & 7.2 & 0.83 & 52.10 & -2342.11$\pm$ 0.43 \\
% Mode6 & 1.39 & 7.5 & 7.5 & 0.84 & 54.12 & -2351.93 $\pm$ 0.68 \\
% Mode7 & 1.22 & 7.5 & 7.5 & 0.85 & 38.72 & -2355.02 $\pm$ 0.14 \\
% Mode8 & 1.38 & 7.3 & 7.4 & 0.82 & 55.47 & -2357.56 $\pm$ 0.86 \\
% Mode9 & 1.32 & 8.2 & 8.2 & 0.86 & 20.21 & -2454.75 $\pm$ 0.87 \\

% \hline
% \end{tabular}
% }

% \caption{Maximum A Posteriori values for each mode (ordered by evidence) for each \texttt{Multinest} live points choices.}
%     \label{tab:live_points}
% \end{minipage}
% \end{table*}

% \usepackage{multirow}
\begin{table*}
\centering
 \begin{minipage}{2.\columnwidth}
    \resizebox{\columnwidth}{!}{%
    
\begin{tabular}{cccccccc}
\hline \hline
\multicolumn{2}{l}{Parameter}                    & $M$ (\msol) & $R_\mathrm{eq}$ (km) & $D$ (kpc) & $\cos(i)$ & $N_H$ ($10^{20}\mathrm{cm}^{-2}$) & Log-Evidence \\

\hline

\multicolumn{1}{c}{\multirow{3}{*}{$2\times10^3$ live points}} & Mode1 & 1.23 & 7.3 & 7.3 & 0.85 & 27.79 & -2256.57$\pm$ 0.09 \\
\multicolumn{1}{c}{}                    &  Mode2 & 1.25 & 7.9 & 7.8 & 0.84 & 42.53 & -2308.18 $\pm$ 0.94 \\
\multicolumn{1}{c}{}                    &  Mode3 & 1.42 & 7.7 & 7.6 & 0.83 & 53.69 & -2314.70 $\pm$ 0.70 \\

\hline

\multirow{9}{*}{$10^5$ live points}     &  Mode1 & 1.23 & 7.4 & 7.6 & 0.85 & 26.14 & -2258.19 $\pm$ 0.12 \\
                                        &  Mode2 & 1.31 & 8.2 & 8.0 & 0.85 & 45.69 & -2309.25$\pm$ 0.13 \\
                                        &  Mode3 & 1.32 & 7.1 & 7.1 & 0.82 & 56.15 & -2314.86 $\pm$ 0.10 \\
                                        &  Mode4 & 1.29 & 8.0 & 7.9 & 0.86 & 46.74 & -2336.74$\pm$ 0.56 \\
                                        &  Mode5 & 1.35 & 7.2 & 7.2 & 0.83 & 52.10 & -2342.11$\pm$ 0.43 \\
                                        &  Mode6 & 1.39 & 7.5 & 7.5 & 0.84 & 54.12 & -2351.93 $\pm$ 0.68 \\
                                        &  Mode7 & 1.22 & 7.5 & 7.5 & 0.85 & 38.72 & -2355.02 $\pm$ 0.14 \\
                                        &  Mode8 & 1.38 & 7.3 & 7.4 & 0.82 & 55.47 & -2357.56 $\pm$ 0.86 \\
                                        & Mode9 & 1.32 & 8.2 & 8.2 & 0.86 & 20.21 & -2454.75 $\pm$ 0.87 \\

\hline
\end{tabular}
}

\caption{Maximum A Posteriori values for each mode (ordered by evidence) for each \texttt{Multinest} live points choices.}
    \label{tab:live_points}
\end{minipage}
\end{table*}